\documentclass[envcountsect]{llncs} 
\usepackage{etex} 
\usepackage[]{graphicx}
\newif\ifignore 
\ignorefalse
\newcommand{\auxproof}[1]{
\ifignore\mbox{}\newline
\textbf{BEGIN: AUX-PROOF} \dotfill\newline
{#1}\mbox{}\newline
\textbf{END: AUX-PROOF}\dotfill\newline
\fi}

\usepackage{times}

\usepackage{theorem}
\usepackage{fancybox,amssymb,amstext,amsmath,stmaryrd,wasysym,cite,proof,mathtools}
\usepackage[pdf]{pstricks}
\usepackage[pdftex,all]{xy}
\usepackage{xspace}
\allowdisplaybreaks[1] 

\usepackage{algorithm}
\usepackage{algorithmic}
\usepackage{here}
\usepackage{bussproofs}
\usepackage{braket}
\usepackage{listings}
\usepackage{url}

\renewcommand{\tilde}{\widetilde}

\xyoption{v2}
\xyoption{curve}
\xyoption{2cell}
\SelectTips{cm}{}  
\UseAllTwocells
\SilentMatrices
\def\labelstyle{\scriptstyle}
\def\twocellstyle{\scriptstyle}
\newdir{ >}{{}*!/-8pt/@{>}}
\newdir{|>}{%
!/1.6pt/@{|}*:(1,-.2)@^{>}*:(1,+.2)@_{>}}
\newdir{pb}{:(1,-1)@^{|-}}
  \def\pb#1{\save[]+<20 pt,0 pt>:a(#1)\ar@{pb{}}[]\restore}

\usepackage{wrapfig}
\theorembodyfont{\itshape}
\newtheorem{mytheorem}{Theorem}[section]
\newtheorem{mylemma}[mytheorem]{Lemma}
\newtheorem{myproposition}[mytheorem]{Proposition}

\theorembodyfont{\rmfamily}

\newtheorem{myremark}[mytheorem]{Remark}
\newtheorem{myexample}[mytheorem]{Example}
\newtheorem{mydefinition}[mytheorem]{Definition}

\spnewtheorem*{myproof}{Proof}{\itshape}{\rmfamily}

\def\myqed{\qed}

\newcommand{\Kleisli}[1]{\mathcal{K}{\kern-.2ex}\ell(#1)}
\newcommand{\EM}[1]{\mathcal{E}{\kern-.5ex}\mathcal{M}(#1)}

\newcommand{\sem}[1]{\llbracket #1 \rrbracket}

\newcommand{\bbP}{\mathbb{P}}

\newcommand{\place}{\underline{\phantom{n}}\,} 

\newcommand{\nat}{\mathbb{N}}
\newcommand{\real}{\mathbb{R}}

\newcommand{\textlb}{
  \def\labelstyle{\textstyle}
  \def\twocellstyle{\textstyle}}

\def\compsign{\mathrel>\kern-2pt\joinrel>\kern-2pt\joinrel>}

\newcommand{\dar}{\ar@{..>}}
\newcommand{\lar}{\ar@{-}}

\newcommand{\ttrue}{\mathtt{t{\kern-1.5pt}t}}
\newcommand{\ffalse}{\mathtt{f{\kern-1.5pt}f}}

\newcommand{\ssub}{\mathrel{\subset{\kern-1.6ex}\subset}}

\newcommand{\PredKl}[1]{\bbP^{\mathcal{K}{\kern-.2ex}\ell}(#1)}
\newcommand{\PredEM}[1]{\bbP^{\mathcal{E}{\kern-.5ex}\mathcal{M}}(#1)}

\newcommand{\UP}{\mathcal{U{\kern-.3ex}P}}
\newcommand{\CD}{\mathcal{C{\kern-.3ex}D}}

\newcommand{\RC}{\mathcal{C}{\kern-.3ex}v}

\newcommand{\cone}{\mathcal{C}}
\newcommand{\multMon}{\mathcal{M}}
\newcommand{\ideal}{\mathcal{I}}
\newcommand{\scone}{\mathcal{SC}}

\newcommand{\FindSOS}{\mathrm{FindSOS}}
\newcommand{\FindPoly}{\mathrm{PDioph}^{\mathrm{SOS}}}
\newcommand{\NULL}{\mathrm{FAIL}}
\newcommand{\diag}[1]{\mathrm{diag}}
\newcommand{\AISAT}{\mathtt{aiSat}}
\newcommand\todo[1]{{\color{red}\bf TODO: {#1}}}
\newcommand{\SASdiseq}{SAS${}_{\neq}$}
\newcommand{\SASstrineq}{$\text{SAS}_{<}$}
\newcommand{\banme}[2]{\stackrel{#1}{\stackrel{\vee}{#2}}}
\newcommand{\SSPolyInt}{\textsc{SSInt}}
\newcommand{\CFE}{\mathsf{CFE}}
\def\middot{\textperiodcentered~}

 \title{Sharper and Simpler Nonlinear Interpolants\\ for Program Verification}
 \author{
 Takamasa Okudono\inst{1}
 \and
 Yuki Nishida\inst{2}
 \and
 Kensuke Kojima\inst{2}
 \and\\
 Kohei Suenaga\inst{2,3}
 \and
 Kengo Kido\inst{1,4}
 \and
 Ichiro Hasuo\inst{5}
 }
 \institute{
   University of Tokyo, Tokyo, Japan\\
   \email{
tokudono@is.s.u-tokyo.ac.jp
  }
    \and
    Kyoto University, Kyoto, Japan\\
    \and
    JST PRESTO, Kyoto, Japan
    \and
    JSPS Research Fellow, Tokyo, Japan
    \and
    National Institute of Informatics, Tokyo, Japan\\
}

\begin{document}

\maketitle

 \begin{abstract}
 \emph{Interpolation} of jointly infeasible predicates plays important roles in various program verification techniques such as invariant synthesis and CEGAR. Intrigued by the recent result by Dai et al.\ that combines real algebraic geometry and SDP optimization in synthesis of polynomial interpolants, the current paper contributes its enhancement that yields  \emph{sharper} and \emph{simpler} interpolants. The enhancement is made possible by: theoretical observations in real algebraic geometry; and our continued fraction-based algorithm that rounds off (potentially erroneous) numerical solutions of SDP solvers.
Experiment results support our tool's effectiveness; we also demonstrate the benefit of sharp and simple interpolants in program verification examples. 

 \end{abstract}

\section{Introduction}\label{sec:intro}
\textbf{Interpolation for Program Verification}\quad
\emph{Interpolation} in logic is a classic problem. Given formulas
$\varphi$ and $\psi$ that are jointly unsatisfiable (meaning 
$\models \varphi\land \psi\Rightarrow\bot$), one asks for a ``simple'' formula
$\xi$ such that 
$\models \varphi\Rightarrow\xi$ and 
$\models \xi\land \psi\Rightarrow\bot$.
The simplicity requirement on $\xi$ can be a formal one (like the \emph{common variable condition}, see Def.~\ref{def:interpolant}) but it can also be informal, like ``$\xi$ is desirably much simpler than $\varphi$ and $\psi$ (that are gigantic).'' Anyway the intention is that $\xi$ should be a simple witness for the joint unsatisfiability of $\varphi$ and $\psi$, that is, an ``essential reason'' why $\varphi$ and $\psi$ cannot coexist. 

This classic problem of interpolation has found various applications in
static analysis and
program
verification~\cite{McMillan03,GurfinkelRS13,JhalaM05,HenzingerJMM04,McMillan2006,McMillan2005}.
This is  particularly  the case with techniques based on automated reasoning, where one relies on symbolic \emph{predicate abstraction} in order to  deal with infinite-state systems like (behaviors of) programs. 
It is crucial for the success of such techniques that we discover ``good'' predicates that capture the essence of the systems' properties. Interpolants---as simple witnesses of incompatibility---have proved to be potent candidates for these ``good'' predicates. 

\vspace{.7em}
\noindent
\textbf{Interpolation  via Optimization and Real Algebraic Geometry}\quad
A lot of research efforts have been made towards efficient interpolation algorithms. One of the earliest is~\cite{ColonSS03}: it relies on \emph{Farkas' lemma} for synthesizing linear interpolants (i.e.\ interpolants expressed by linear inequalities). This work and  subsequent ones
have signified the roles of \emph{optimization} problems and their algorithms in efficient  synthesis of interpolants. 

In this line of research we find the recent contribution by Dai et al.~\cite{DaiXZ13} remarkable, from both theoretical and implementation viewpoints. Towards synthesis of \emph{nonlinear} interpolants (that are expressed by polynomial inequalities), their framework in~\cite{DaiXZ13} works as follows. 
\begin{itemize}
 \item On the theory side it relies on \emph{Stengle's Positivstellensatz}---a fundamental result in \emph{real algebraic geometry}~\cite{Stengle74,BochnakCR99}---and relaxes the interpolation problem to the problem of finding a suitable ``disjointness certificate.'' The latter consists of a few polynomials subject to certain conditions. 
 \item On the implementation side it relies on state-of-the-art \emph{SDP solvers} to efficiently solve the SDP problem that results from the above relaxation. 
\end{itemize}
In~\cite{DaiXZ13} it is reported that the above framework successfully synthesizes nontrivial nonlinear interpolants, where some examples are taken from program verification scenarios. 

\vspace{.7em}
\noindent
\textbf{Contribution}\quad
The current work contributes an enhancement of the  framework from~\cite{DaiXZ13}. Our specific concerns are in \emph{sharpness} and \emph{simplicity} of interpolants.

\vspace{.7em}
\noindent
\begin{minipage}{.8\textwidth}
  \begin{myexample}[sharp interpolant]\label{ex:sharpInterpolantIntro}
 Let 
 \begin{math}
  \mathcal{T} \coloneqq (y>x \land x>-y)
\end{math} and
 \begin{math}
 \mathcal{T}' \coloneqq (y\le -x^2)
 \end{math}. These designate the blue and red areas in the figure,
   respectively.  We would like an interpolant $\mathcal{S}$ so that
   $\mathcal{T}$ implies $\mathcal{S}$ and $\mathcal{S}$ is disjoint
   from  $\mathcal{T}'$. Note however that such an interpolant
   $\mathcal{S}$  must be ``sharp.'' The areas of $\mathcal{T}$ and
   $\mathcal{T}'$ almost intersect with each other at $(x,y)=(0,0)$.
   That is,
   the conditions $\mathcal{T}$ and    $\mathcal{T}'$ are barely
   disjoint in the sense that, once we replace $>$ with $\ge$ in
   $\mathcal{T}$, they are no longer disjoint. (See
   Def.~\ref{def:symbolicclosure} for formal definitions.)
       \end{myexample}
\end{minipage}
\begin{minipage}{.19\textwidth}
  \includegraphics[clip,trim=0 0 0cm
  0,width=\textwidth]{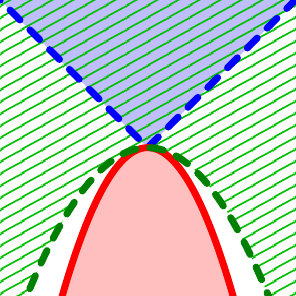}
\end{minipage}

\vspace*{.1em}

The original  framework in~\cite{DaiXZ13} fails to synthesize such ``sharp'' interpolants; and this failure is theoretically guaranteed (see \S{}\ref{subsec:limitationOfDai}). In contrast our modification of the framework succeeds: it yields an interpolant  $8y+4x^2 > 0$ (the green hatched area). 

\vspace*{.5em}
\noindent
\begin{minipage}{.8\textwidth}
 \begin{myexample}[simple interpolant]\label{ex:notSimpleInterpolant}
 Let
 \begin{math}
 \mathcal{T} \coloneqq (y\ge x^2 + 1)
 \end{math} 
 and
 \begin{math}
 \mathcal{T}' \coloneqq (y\le -x^2 -1)
 \end{math}.
The implementation $\AISAT$~\cite{aiSatGitHub20170117} of the workflow in~\cite{DaiXZ13} succeeds and synthesizes an interpolant $284.3340y +0.0012x^2y > 0$. In contrast our tool synthesizes $5y+2>0$ that is much simpler.
 \end{myexample}
\end{minipage}
\begin{minipage}{.19\textwidth}
  \includegraphics[width=\textwidth]{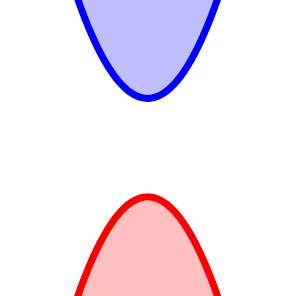}
\end{minipage}

\vspace*{.5em}
The last two examples demonstrate  two issues that we found in the original framework in~\cite{DaiXZ13}.
 Our enhanced framework shall address these issues of sharpness and simplicity, employing the following two main technical pieces.

 The first piece is \emph{sharpened Positivstellensatz-inspired
 relaxation} (\S{}\ref{sec:limitationOfDai}).   We start with the relaxation
 in~\cite{DaiXZ13} that reduces interpolation to finding polynomial
 certificates. We devise its ``sharp'' variant that features: use of \emph{strict inequalities} $>$ (instead of \emph{disequalities} $\neq$); and a corresponding adaptation of Positivstellensatz that uses a notion we call \emph{strict cone}. Our sharpened relaxation allows encoding to SDP problems, much like in~\cite{DaiXZ13}. 

The second technical piece that we rely on is our continued fraction-based \emph{rounding algorithm}. We employ the algorithm in what we call the \emph{rounding-validation loop} (see~\S{}\ref{sec:impl}), a workflow from~\cite{Harrison2007} that addresses
 the challenge of \emph{numerical errors}. 

Numerical relaxation of problems in automated reasoning---such as the SDP relaxation in~\cite{DaiXZ13} and in the current work---is nowadays common, because of potential performance improvement brought by numerical solvers. However a numerical solution is subject to
 numerical errors, and due to those errors, the solution may not satisfy the original constraint. This  challenge is identified by many authors~\cite{RouxVS16,Harrison2007,Besson2007,PEYRL2008269,KaltofenLYZ08,PlatzerQR09}. 

Moreover, even if a numerical solution satisfies the original
constraint, the solution often involves floating-point numbers and thus
is not  simple. See Example~\ref{ex:notSimpleInterpolant}, where one may
wonder if the coefficient $0.0012$ should be simply $0$. Such
complication is a disadvantage in applications in program verification,
where we use interpolants as candidates for ``useful'' predicates. These
predicates should grasp essence and reflect insights of programmers;  it
is our  hypothesis  that such predicates should be simple. Similar
arguments have been made in previous works such as~\cite{JhalaM06,Terauchi15}. 

 To cope with the last challenges of potential unsoundness and lack of simplicity, we employ a workflow that we call the \emph{rounding-validation loop}. The workflow has been used e.g.\ in~\cite{Harrison2007}; see Fig.~\ref{fig:overviewOfThetool} (pp.~\pageref{fig:overviewOfThetool}) for a schematic overview. In the ``rounding'' phase we apply our continued fraction-based rounding algorithm to a candidate obtained as a numerical solution of an SDP solver.
In the ``validation'' phase the rounded candidate is fed back to the original constraints and their satisfaction is checked by purely symbolic means. If validation fails, we increment the \emph{depth} of rounding---so that the candidate becomes less simple but closer to the original candidate---and we run the loop again.

\vspace{.7em}
\noindent
\begin{minipage}{.84\textwidth}
 \begin{myexample}[invalid interpolant candicate]\label{ex:invalidCandidateIntro}
Let \begin{math}
  \mathcal{T} = (y \le -1),
 \mathcal{T}' = (x^2+y^2 < 1)
\end{math}, as shown in the figure.  These are barely disjoint and hence the algorithm in~\cite{DaiXZ13} does not apply to it.
In our workflow, the first interpolant candidate that an SDP solver yields is $f(x,y)\ge 0$, where 
\end{myexample}
\end{minipage}
\hfill
\begin{minipage}{.14\textwidth}\centering
  \includegraphics[width=\textwidth]{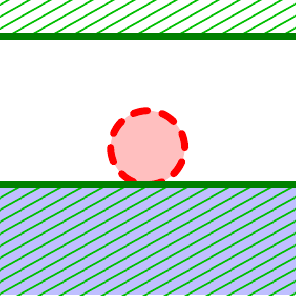}
\end{minipage}

\begin{displaymath}
f(x,y) =
 \left(
 \begin{minipage}{.75\textwidth}\scriptsize
 $  -3.370437975+8.1145\times 10^{-14}x-2.2469y+1.1235y^2-2.2607\times 10^{-10}y^3
 +9.5379\times 10^{-11}x^2-2.2607\times 10^{-10}x^2 y -4.8497\times 10^{-11}x^2y^2-1.1519\times 10^{-14}x^3
 +4.8935\times 10^{-11}x^4-9.7433\times 10^{-11}y^4 $
 \end{minipage}
\right)\enspace. 
\end{displaymath}
Being the output of a numerical solver the coefficients are far from simple integers. Here coefficients in very different scales coexist---for example one may wonder if the coefficient $8.1145\times 10^{-14}$ for $x$ could just have been $0$. Worse,  the above candidate is in fact not an interpolant: $x=0,y=-1$ is in the region of $\mathcal{T}$ but we have $f(0,-1)<0$. 

By subsequently applying our  rounding-validation loop, we eventually obtain a candidate $34y^2 - 68y - 102 \ge 0$, and its validity is guaranteed by our tool.

This workflow of the rounding-validation loop is adopted from~\cite{Harrison2007}. Our technical contribution lies in the rounding algorithm that we use therein. It can be seen
 an extension of the well-known rounding procedure by \emph{continued fraction expansion}. The original procedure, employed e.g.\ in~\cite{PEYRL2008269}, rounds a real number into a rational number (i.e.\ a ratio $k_{1}\colon k_{2}$ between two integers). In contrast, our current extension rounds a ratio $r_{1}\colon \cdots\colon r_{n}$ between $n$ real numbers into a simpler ratio $k_{1}\colon \cdots\colon k_{n}$.

We have implemented our enhancement of~\cite{DaiXZ13}; we call our tool
\SSPolyInt\ (\emph{Sharp and Simple Interpolants}). Our experiment
results  support its effectiveness: the tool succeeds in synthesizing
sharp interpolants (while the workflow in~\cite{DaiXZ13} is guaranteed
to fail); and our program verification examples demonstrate the benefit
of sharp and simple interpolants (synthesized by our tool)  in
verification. The latter benefit is demonstrated by the following
example; the example is discussed in further detail later in~\S{}\ref{sec:experiments}.

  \begin{myexample}[program verification]\label{ex:programverification}
Consider the imperative program in Listing~\ref{codeDai11} (pp.~\pageref{codeDai11}). Let us verify its assertion (the last line) by \emph{counterexample-guided abstraction
refinement (CEGAR)}~\cite{DBLP:journals/jacm/ClarkeGJLV03}, in which we
   try to synthesize suitable predicates that separate the reachable
   region (that is under-approximated by finitely many samples of execution traces) and the unsafe region ($(xa) + 2 (ya) < 0$). The use of interpolants as candidates for such separating predicates has been advocated by many authors, including~\cite{HenzingerJMM04}. 

Let us say that the first    
   execution trace we sampled is the
   one in which the while loop is not executed at all ($1\rightarrow 2 \rightarrow 3 \rightarrow 4 \rightarrow
16$ in line numbers). Following the workflow of CEGAR by interpolation,
   we are now required to compute an interpolant of
$\mathcal{T} := (xa=0 \land
ya=0)$ and $\mathcal{T'} := ((xa) + 2 (ya) < 0)$.
Because $\mathcal T$ and $\mathcal T'$ are ``barely disjoint'' (in the sense of Example~\ref{ex:sharpInterpolantIntro}, that is, the closures of $\mathcal{T}$ and $\mathcal{T'}$ are no longer disjoint),
the  procedure in~\cite{DaiXZ13} cannot generate any interpolant. In
   contrast, our implementation---based on our refined use of Stengle's
   positivstellensatz, see~\S{}\ref{sec:limitationOfDai}---successfully discovers an interpolant
   $(xa)+2 (ya) \ge 0$. This interpolant happens to be an invariant of
   the program and proves the safety of the program.

   Later in~\S{}\ref{sec:experiments} we explain this example in further
   detail. 

   \end{myexample}

\vspace{.5em}
\noindent
\textbf{Related Work}\quad
Aside from
the work by Dai et al.~\cite{DaiXZ13} on which we are based, 
there are
several approaches to polynomial interpolation in the literature.  Gan
et al.~\cite{GanDXZKC16} consider interpolation for
polynomial inequalities that involve uninterpreted functions, with
the restriction that the degree of polynomials is quadratic. An earlier work with a similar aim is~\cite{KupferschmidB11} by Kupferschmid et al. 
Gao and Zufferey~\cite{GaoZ16} study nonlinear interpolant synthesis
over real numbers.  Their method can handle transcendental functions
as well as polynomials.  Interpolants are generated from refutation,
and represented as union of rectangular regions.
Because of this representation, although their method enjoys $\delta$-completeness (a notion of approximate   completeness), it cannot synthesize sharp interpolants
like Example~\ref{ex:sharpInterpolantIntro}.
Their interpolants
tend to be fairly
complicated formulas, too, and therefore would not necessarily be suitable for
applications like
program
verification (where we seek simple predicates; see \S{}\ref{sec:experiments}).

\emph{Putinar's positivstellensatz}~\cite{Putinar93} is a well-known variation of Stengle's positivstellensatz; it is known to allow simpler SDP relaxation than Stengle's. However it does not suit the purpose of the current paper because: 1) it does not allow mixture of strict and non-strict inequalities; and 2) it requires a compactness condition. There is a common trick to force strict inequalities in a framework that only allows non-strict inequalities, namely to add a small perturbation. We find that this trick does not work in our program verification examples; see~\S{}\ref{sec:experiments}. 

The problem with numerical errors in SDP solving has been discussed in
the literature.  Harrison~\cite{Harrison2007} is one of the first to
tackle the problem: the work introduces the workflow of the
rounding-validation loop; the rounding algorithm used there increments a denominator at each step and thus is simpler
than our continued fraction-based one. The same rounding algorithm is used in~\cite{Besson2007}, as we observe in the code. 
Peyrl \& Parrilo~\cite{PEYRL2008269}, towards the goal of sum-of-square decomposition in rational coefficients, employs a rounding algorithm by continued fractions. The difference from our current algorithm is that they apply continued fraction expansion to each of the coefficients, while our generalized algorithm simplifies the ratio between the coefficients altogether. The main technical novelty of~\cite{PEYRL2008269} lies in identification of a condition for validity of a rounded candidate. This framework is further extended in Kaltofen et al.~\cite{KaltofenLYZ08} for a different optimization problem, combined with the  Gauss--Newton iteration. 

More recently, 
an approach using a simultaneous Diophantine approximation
algorithm---that computes the best approximation within a given bound
of denominator---is considered by Lin et al.~\cite{Lin2014}.
They focus on finding a rational fine approximation to the output of
SDP solvers, and do not aim at simpler certificates.
Roux et al.~\cite{RouxVS16} proposes methods that guarantee existence of a solution
relying on numerical solutions of SDP solvers.  They mainly focus on strictly feasible
problems, and therefore some of our examples in~\S{}\ref{sec:experiments} are out of their scope.
Dai et al.~\cite{DaiGXZ17} address
the same problem of numerical errors
 in the context of barrier-certificate synthesis.
They use \emph{quantifier elimination} (QE)
for validation,
while our validation method  relies
 on a well-known
characterization of positive semidefiniteness
(whose check is less expensive than QE; see~\S{}\ref{subsec:validation}).

\auxproof{
(*** Takamasa, add a paragraph!! ***)
We have to obtain the exact certificate to use semidefinite programming for verification, but numerical error is inevitable.  There are some attempts to get certificates from the numerical result of semidefinite programming.
\begin{itemize}
 \item Harrison~\cite{Harrison2007} designed an algorithm for the automated verification of nonlinear formulas exploiting semidefinite programming and implemented it in HOL Light.

The main trick is rounding the numerical result by a rational number whose denominator is $d$ for each elements.  If the rounded result is exactly an positive semidefinite, then the certificate is generated.  If it fails, then increase $d$ by one and retry.  

The author says the method exactly works well.  He applied it to the problem whose matrix size is $38$ parametrized by 143 variables and the verification succeeded.

While the loop of checking positive semidefiniteness and increasing the denominator is natural, it seems that there are redundant steps.  For example, let us think about rounding $3.14$.  The 7th rounding is $22/7$ and the 8th is $25/8$.  The gap between the 8th and the original value $3.14$ is bigger than that of 7th, so the increment does not contribute to the improvement of the accuracy.  Such an redundancy is studied in the field of approximation of a real number, and it is known as the theory of Diophantine approximation~\cite{lang1995Dioph}.  We designed an approximation algorithm avoiding such an redundancy and presented it in section \ref{subsec:rounding}.

 \item 
Besson~\cite{Besson2007} implemented a Coq tactic \texttt{micromega} that supports nonlinear goals.  The approach is different from Harrison~\cite{Harrison2007}, but the approximation method is essentially similar according to the implementation~\cite{githubcoq170427}.
 \item 
Peyrl~\cite{PEYRL2008269} proposed an algorithm for sum of square decomposition in rational coefficients, that is, a method to represent a multivariate polynomial by sums of polynomials whose coefficients are rational.  

The approximation method is to apply Diophantine approximations for each elements in the result matrix of the SDP solver.  Moreover, Peyrl gives the criteria of the sufficient depth of the approximation by observing the constraints of the SDP problem.  The system of linear equations in the SDP problem defines an affine space in the space of symmetric matrices, so it is possible to project the result of SDP onto the affine space from the space of matrices.  A theorem in the paper gives the way to calculate the tolerance of approximation from the length of the perpendicular of the projection and the minimum eigenvalue of the result of the SDP.  Of course, there must be some kind of robustness to apply this method because the approximation causes an perturbation to the result of SDP.  The robustness means that all the eigenvalues in the result of SDP must be strictly positive.  The condition of the robustness is called strictly feasible.  
The assumption of strict feasibility is not always satisfied, but the approximation itself works well experimentally.  
 \item 
Kaltofen et al.\cite{KaltofenLYZ08} combined Peyrl's work and the numerical optimization method Gauss-Newton iteration and got more accuracy.  After the iteration, the projection onto the the affine space is done by solving least squares problem.  It is possible to solve the problem in rational numbers, so the projection can be used as the rational approximation.
 \item \cite{JFR4319}, Magron: I cannot understand...
 \item \cite{Parrilo03}, Parrilo
\end{itemize}
}

\vspace{.5em}
\noindent
\textbf{Future Work}\quad
The workflow of the rounding-validation loop~\cite{Harrison2007} is simple but potentially effective:
in combination with our rounding algorithm based on continued fractions,  we speculate that the workflow can offer a general methodology for coping with numerical errors in verification and in symbolic reasoning. 
Certainly our current implementation is not the best of the
workflow: for example, 
the validation phase of~\S{}\ref{sec:impl} could be further improved by
techniques from interval arithmetic, e.g.\ from~\cite{Rump2006}. 
 

 Collaboration between  numerical and symbolic computation in general (like in~\cite{AnaiP03}) interests us, too. For example in our workflow (Fig.~\ref{fig:overviewOfThetool}, pp.~\pageref{fig:overviewOfThetool}) there is a disconnection between the SDP phase and later: passing additional  information (such as gradients) from the SDP phase can make the rounding-validation loop more effective. 


Our current examples  are rather simple and small. While
they serve as a feasibility study of the proposed interpolation method,
practical applicability of the method in the context of program
verification is yet to be confirmed. We plan to conduct more extensive
case studies, using common program verification benchmarks such as 
in~\cite{Sharma2013}, making comparison with other methods, and further
refining our method in its course.

\vspace{.5em}
\noindent
\textbf{Organization of the Paper}\quad
In~\S{}\ref{sec:prelim} we review the framework in~\cite{DaiXZ13}. Its lack of sharpness is established in~\S{}\ref{subsec:limitationOfDai}; this motivates our sharpened Positivstellensatz-inspired relaxation of interpolation in~\S{}\ref{subsec:sharperInterpolationViaPositivstellensatz}. In~\S{}\ref{sec:impl} we describe our whole workflow and its implementation, describing the rounding-validation loop and the continued fraction-based algorithm used therein.
In~\S{}\ref{sec:experiments} we present experimental results and discuss the benefits in program verification.
Some proofs and details are deferred to appendices.

\section{Preliminaries}\label{sec:prelim}
Here we review the previous interpolation algorithm by Dai et al.~\cite{DaiXZ13}. It is preceded by its mathematical bedrock, namely  Stengle's Positivstellensatz~\cite{Stengle74}. 
\subsection{Real Algebraic Geometry and Stengle's Positivstellensatz}
\label{subsec:RAC}

We write $\vec{X}$ for a sequence $X_1,X_2,\dots,X_k$ of variables, and
$\real[\vec X]$ for the set of polynomials in $X_1,  \dots, X_k$
over $\real$. We sometimes write $f(\vec X)$ for a polynomial $f\in\real[\vec X]$ in order to signify that the variables  in $f$ are restricted to those  in $\vec X$.

\begin{mydefinition}[\SASdiseq]\label{def:SAS}
A \emph{semialgebraic system with disequalities (\SASdiseq)} $\mathcal{T}$, in variables  $X_1,X_2,\dots,X_k$,  is
a sequence 
\begin{equation}\label{eq:SAS}\small
 \mathcal{T}
\;=\;
\left(
\begin{array}{l}
   f_{1}(\vec{X}) \ge 0\enspace,
  \;\dotsc,\;
  f_{s}(\vec{X}) \ge 0\enspace,
  \qquad
  g_{1}(\vec{X}) \neq 0\enspace,
  \;\dotsc,\;
  g_{t}(\vec{X}) \neq 0\enspace,
  \\
  h_{1}(\vec{X}) = 0\enspace,
  \;\dotsc,\;
  h_{u}(\vec{X}) = 0
\end{array}
\right)
\end{equation}
 of \emph{inequalities} $f_{i}(\vec{X})\ge 0$, 
\emph{disequalities} $g_{j}(\vec{X})\neq 0$ and
\emph{equalities} $h_{k}(\vec{X})=0$. 
Here $f_{i}, g_{j}, h_{k}\in \real[\vec{X}]$ are polynomials, for $i\in [1,s], j\in [1,t]$ and $k\in [1,u]$. 
 
For the \SASdiseq\ $\mathcal{T}$ in~(\ref{eq:SAS}) in $k$ variables, we say $\vec x \in \real^k$ \emph{satisfies} $\mathcal{T}$ if 
$f_{i}(\vec{x})\ge 0$, 
$g_{j}(\vec{x})\neq 0$ and
$h_{k}(\vec{x})=0$ hold for all $i,j,k$. We let $\sem{\mathcal{T}}\subseteq\real^{k}$ denote the set of all such $\vec{x}$, that is, 
\begin{math}
\sem{\mathcal{T}} \coloneqq \set{{\vec{x}} \in {\real}^k | \text{$\vec x$ satisfies $\cal T$}} 
\end{math}.
\end{mydefinition}




\begin{mydefinition}[cone, multiplicative monoid, ideal]\label{def:coneMMIdeal}
 A set $C\subseteq\real[\vec X]$ is a \emph{cone} 
if it satisfies the following closure properties:
1) $f,g\in C$ implies $f+g \in C$;
2) $f,g\in C$ implies $fg\in C$; and
3) $f^2 \in C$ for any
$f\in \real[\vec X]$.

  A set $M\subseteq\real[\vec X]$ is a \emph{multiplicative
    monoid} 
if it satisfies the following:
1) $1 \in M$; and
2) $f,g \in M$ implies $fg \in M$.

 A set $I\subseteq\real[\vec X]$ is an \emph{ideal} if it satisfies:
1) $0\in I$;
2) $f,g\in I$ implies $f+g\in I$; and
3) $fg\in I$ for any $f\in \real[\vec X]$ and $g\in I$.

For a subset $A$ of $\real[\vec{X}]$, we write:
$\cone(A)$, $\multMon(A)$, and $\ideal(A)$ for the smallest cone,
multiplicative monoid, and ideal, respectively, that includes $A$.
 \end{mydefinition}



\noindent
The last notions encapsulate closure properties of inequality/disequality/equality predicates, respectively, in the following sense.
 The definition of $\sem{\mathcal{T}}\subseteq\real^{k}$ is in Def.~\ref{def:SAS}. 

\begin{mylemma}
\label{lem:closurePropertiesOfCMI}
 Let $\vec x\in \real^{k}$ and $f_{i}, g_{j}, h_{k}\in \real[\vec X]$.
  \begin{enumerate}
  \item If $\vec{x} \in \sem{f_1 \ge 0,\dots,f_s \ge 0}$, then
    $f(\vec x) \ge 0$ for all $f \in \cone(f_1,\dots,f_s)$.
  \item If $\vec x\in \sem{g_1\neq 0,\dots,g_t\neq 0}$, then
    $g(\vec x)\neq 0$ for all $g\in \multMon(g_1,\dots,g_t)$.
  \item If $\vec x \in \sem{h_1=0,\dots,h_u=0}$, then $h(\vec x)=0$
    for all $h\in \ideal(h_1,\dots,h_u)$.
  \qed
  \end{enumerate}
\end{mylemma}

The following theorem is commonly attributed
to~\cite{Stengle74}. See also~\cite{BochnakCR99}.
\begin{mytheorem}[Stengle's Positivstellensatz]\label{thm:positivStellenSatz}
 Let $\mathcal{T}$ be the \SASdiseq\ in~(\ref{eq:SAS}) (Def.~\ref{def:SAS}).
 It is infeasible (meaning $\sem{\mathcal{T}}=\emptyset$) if and
 only if
 there exist $f\in \cone(f_{1},\dotsc,f_{s})$,
$g\in \multMon(g_{1},\dotsc,g_{t})$ and
$h\in \ideal(h_{1},\dotsc,h_{u})$
such that $f+g^{2}+h= 0$.
\qed
\end{mytheorem}
 The polynomials $f,g,h$ can be seen as an \emph{infeasible certificate} of the \SASdiseq\ $\mathcal{T}$. 
 The ``if'' direction is shown easily: if $\vec{x} \in \sem{\mathcal{T}}$ then we have $f(\vec x)\ge 0$, $g(\vec x)^{2} > 0$ and $h(\vec x)=0$ (by Lem.~\ref{lem:closurePropertiesOfCMI}), leading to a contradiction. The ``only if'' direction
is nontrivial and remarkable;  it is however not used in the algorithm of~\cite{DaiXZ13} nor in this paper. 

\emph{SOS polynomials} play important  roles, both theoretically and in implementation. 
\begin{mydefinition}[sum of squares (SOS)]
 A polynomial is called a
  \emph{sum of squares (SOS)} if it
 can be written in the form
  $p_1^2 + \dots + p_N^2$ (for some polynomials $p_1,\dots,p_N$).  Note that
  $\cone(\emptyset)$ is exactly the set of sums of squares (Def.~\ref{def:coneMMIdeal}).
\end{mydefinition}

\subsection{
The Interpolation Algorithm by
Dai et al.}
\label{subsec:DaiEtAlAlgorithm}

\begin{mydefinition}[interpolant]
\label{def:interpolant}
Let  $\mathcal{T}$ and $\mathcal{T}'$ be \SASdiseq's, in variables $\vec X,\vec Y$ and in $\vec X, \vec Z$, respectively, given in the following form. Here we assume that each variable in $\vec{X}$ occurs both in $\mathcal{T}$ and $\mathcal{T}'$, and that 
$\vec Y \cap \vec Z = \emptyset$.
\begin{align}
\label{eq:T}
\footnotesize
\!\!\!\!\!\!
\begin{array}{l}
   \mathcal{T}
  =
  \left(
  \begin{array}{l}\footnotesize
    f_{1}(\vec{X},\vec Y) \ge 0,
   \;\dotsc,\;
   f_{s}(\vec{X},\vec Y) \ge 0,
   \quad
   g_{1}(\vec{X},\vec Y) \neq 0,
   \;\dotsc,\;
   g_{t}(\vec{X},\vec Y) \neq 0,
   \\
   h_{1}(\vec{X},\vec Y) = 0,
   \;\dotsc,\;
   h_{u}(\vec{X},\vec Y) = 0
  \end{array}
\right)
\\[+.5em]
   \mathcal{T}'
  =
  \left(
  \begin{array}{l}\small
    f'_{1}(\vec{X},\vec Z) \ge 0,
   \;\dotsc,\;
   f'_{s'}(\vec{X},\vec Z) \ge 0,
   \quad
   g'_{1}(\vec{X},\vec Z) \neq 0,
   \;\dotsc,\;
   g'_{t'}(\vec{X},\vec Z) \neq 0,
   \\
   h'_{1}(\vec{X},\vec Z) = 0,
   \;\dotsc,\;
   h'_{u'}(\vec{X},\vec Z) = 0
  \end{array}
  \right)
\end{array}
\end{align}

Assume further that 
$\mathcal{T}$ and $\mathcal{T}'$ are \emph{disjoint}, that is,
  $\sem{\mathcal{T}}\cap \sem{\mathcal{T}'}=\emptyset$.

An \SASdiseq\ $\mathcal{S}$ is an \emph{interpolant} of $\mathcal{T}$ and $\mathcal{T}'$ if
it satisfies the following: 
\begin{enumerate}
 \item 
$\sem{\mathcal{T}} \subseteq \sem{\mathcal{S}}$;
 \item 
$\sem{\mathcal{S}} \cap \sem{\mathcal{T}'} = \emptyset$; and
 \item (the \emph{common variable condition}) the \SASdiseq\ $\mathcal{S}$ is in the variables $\vec{X}$, that is, $\mathcal{S}$ contains only those variables which occur both in $\mathcal{T}$  and $\mathcal{T'}$. 
\end{enumerate}

%

\end{mydefinition}

Towards efficient synthesis of nonlinear interpolants
Dai et al.~\cite{DaiXZ13} 
introduced a workflow that hinges on the following
variation of Positivstellensatz. 
\begin{mytheorem}[{{disjointness certificate in~\cite[\S{}4]{DaiXZ13}}}]
\label{thm:FromCertificateToInterpolant}
 Let $\mathcal{T},\mathcal{T}'$ be the \SASdiseq's in~(\ref{eq:T}).  
 Assume  there exist
 \begin{align}
 \label{eq:1plus}
&\begin{array}{l}
   \tilde f\in \cone(f_1,\dots,f_s,f'_1,\dots,f'_{s'})\enspace,
  \quad
  g\in \multMon(g_1,\dots,g_t,g'_1,\dots,g'_{t'})
  \quad\text{and}\quad
  \\
  \tilde h\in \ideal(h_1,\dots,h_{u},h'_1,\dots,h'_{u'})\enspace,
  \quad\text{such that}\quad 
   1+\tilde f+g^2+\tilde h\;=\;0\enspace.
\end{array} 
\end{align}
Assume further that 
 $\tilde f$ allows a decomposition $\tilde f=f+f'$, with some $f\in \cone(f_1,\dots,f_s)$ and $f'\in \cone(f'_1,\dots,f'_{s'})$. 
(An element $\tilde h$ in the ideal always allows a decomposition  $\tilde h = h+h'$ such that $h\in \ideal(h_1,\dots,h_u)$ and $h'\in \ideal(h'_1,\dots,h'_{u'})$.)

Under the assumptions
 $\mathcal{T}$ and $\mathcal{T}'$ are disjoint. Moreover the \SASdiseq\ 

\noindent
\begin{minipage}{\textwidth}
\begin{equation}\label{eq:daiEtAlInterpolant}
\begin{array}{l}
   \mathcal{S} \;\coloneqq\; \left(\,1/2+f+g^2+h > 0\,\right)
\end{array}
\end{equation}
\end{minipage}

\vspace{.5em}
\noindent
satisfies the conditions of an interpolant of $\mathcal{T}$ and $\mathcal{T}'$ (Def.~\ref{def:interpolant}), except for Cond.~3.\ (the common variable condition). 
\end{mytheorem}
\begin{myproof}
The proof is much like the ``if'' part of Thm.~\ref{thm:positivStellenSatz}. 
It suffices to show that $\mathcal{S}$ is an interpolant; then the disjointness of  $\mathcal{T}$ and $\mathcal{T}'$ follows. 

To see $\sem{\mathcal{T}} \subseteq \sem{\mathcal{S}}$, assume $\vec{x}\in \sem{\mathcal{T}}$. Then 
we have $f(\vec x)\ge 0$ and 
$h(\vec x)= 0$ by Lem.~\ref{lem:closurePropertiesOfCMI}; additionally $\bigl(g(\vec x)\bigr)^{2}\ge 0$ holds too. Thus $1/2+f(\vec x)+\bigl(g(\vec x)\bigr)^{2}+h(\vec x)\ge 1/2 >0$ and we have $\vec{x}\in\sem{\mathcal{S}}$. 

To see $\sem{\mathcal{S}}\cap\sem{\mathcal{T}'}=\emptyset$, we firstly observe that the following holds for any $\vec x$. 
\begin{minipage}{\textwidth}
\vspace{-.8em}
\begin{align}\label{eq:propFromCertificateToInterpolant}
\begin{array}{rl}
  0
& =\; 1+f(\vec x) + f'(\vec x)+
 \bigl(g(\vec x)\bigr)^{2}+ h(\vec x)  + h'(\vec x)
 \quad\text{by~(\ref{eq:1plus})
}
\\
& =\;
 \bigl(\,
 1/2 + f(\vec x) + 
 (g(\vec x))^{2}+ h(\vec x) 
\,\bigr)
 +
 \bigl(\,
 1/2 + f'(\vec x)+
  h'(\vec x)
\,\bigr)\enspace.
\end{array}
\end{align}
\end{minipage}

\vspace{.5em}
 Assume $\vec x\in\sem{\mathcal{S}}\cap\sem{\mathcal{T}'}$. 
By $\vec x\in\sem{\mathcal{S}}$ we have $ 1/2 + f(\vec x) + 
 (g(\vec x))^{2}+ h(\vec x)  > 0
$; and by $\vec x\in\sem{\mathcal{T}'}$ we have 
$f'(\vec x)\ge 0$ and 
$h'(\vec x)= 0$ (Lem.~\ref{lem:closurePropertiesOfCMI}), hence $ 1/2 + f'(\vec x)+
  h'(\vec x)
\ge 1/2 > 0$. Thus the right-hand side of~(\ref{eq:propFromCertificateToInterpolant}) is strictly positive,  a contradiction.
\myqed
\end{myproof}
Note that we no longer have completeness: existence of an interpolant like~(\ref{eq:daiEtAlInterpolant}) is not guaranteed. Nevertheless Thm.~\ref{thm:FromCertificateToInterpolant} offers a sound method to construct an interpolant, namely by finding a suitable disjointness certificate $f,f',g, h, h'$. 

\begin{algorithm}[tbp]
  \caption{The interpolation algorithm by Dai et al.~\cite{DaiXZ13}. Here $\mathbf 2=\{0,1\}$}
  \label{alg:Dai}
  \begin{algorithmic}[1]
    \STATE \textbf{input}: \SASdiseq's $\mathcal{T},\mathcal{T}'$ in~(\ref{eq:T}), and $b\in \nat$ (the maximum degree)
    \STATE \textbf{output}: either an interpolant $\mathcal{S}$ of $\mathcal{T}$ and $\mathcal{T}'$, or $\NULL$
    \STATE $h\coloneqq (\prod_{i=1}^t g_i) (\prod_{i'=1}^{t'}{g'_{i'}})
        \; ;\quad
   g\coloneqq h^{\lfloor b/2\deg(h) \rfloor}
   $\qquad
    \COMMENT{$g$ is roughly of degree $b/2$}
    \STATE \label{line:algDaiPDioph}
     Solve $\FindPoly$ to find 
   $(
   \overrightarrow{\alpha},
   \overrightarrow{\alpha'},
   \overrightarrow{\beta},
   \overrightarrow{\beta'}
)$. Here:
\vspace{-1em}
\begin{itemize}
 \item $\alpha_{i} \in \cone(\emptyset)_{\le b}$ (for $i\in \mathbf{2}^s$) and
     $\alpha'_{i'} \in \cone(\emptyset)_{\le b}$ (for $i'\in \mathbf{2}^{s'}$) are SOSs, 
 \item $\beta_{j}\in \real[\vec X]_{\le b}$ (for $j\in [1,u]$) and
     $\beta'_{j'}\in \real[\vec X]_{\le b}$ (for $j'\in [1,u']$) are polynomials, 
 \item and they are subject to the constraint
 \vspace{-1.5em}
       \begin{align}
 1+\sum_{i\in \mathbf{2}^s} \alpha_{i} f_1^{i_1}\dotsm f_s^{i_s} 
+ \sum_{i'\in \mathbf{2}^{s'}} \alpha'_{i'} {f'}_1^{i'_1}\dotsm {f'}_{s'}^{i'_{s'}}
+ g^{2}
+ \sum_{j=1}^u \beta_j h_j + \sum_{j'=1}^{u'}\beta'_{j'} h'_{j'} = 0\enspace.
\label{eq:Dai}   
       \end{align}
\end{itemize}
\vspace{-2em}
(Such 
   $(
   \overrightarrow{\alpha},
   \overrightarrow{\alpha'},
   \overrightarrow{\beta},
   \overrightarrow{\beta'}
)$ may not be found, 
in which case return $\NULL$)
   \STATE\label{line:algDaiDefFH}
    $f\coloneqq \sum_{i\in \mathbf{2}^s} \alpha_{i} f_1^{i_1}\dotsm f_s^{i_s} $
    \;;\quad
   $h\coloneqq \sum_{j=1}^u \beta_j h_j$
   \RETURN $\mathcal{S} \coloneqq \left(1/2+
    f +g^{2} + h > 0\right)$
   \label{line:algDaiReturn}
  \end{algorithmic}
\end{algorithm}

The interpolation algorithm in~\cite{DaiXZ13} is shown in Algorithm~\ref{alg:Dai}, where search for a disjointness certificate $f,f',g, h, h'$ is relaxed to the following problem. 
\begin{mydefinition}[$\FindPoly$]\label{def:findpoly}
Let $\FindPoly$ stand for the following problem.
\label{def:findPoly}
\begin{displaymath}
 \begin{tabular}{ll}
  \text{\bfseries Input: \quad}
 &
 \text{polynomials $\varphi_{1},\dotsc,\varphi_{n},\; \psi_{1},\dotsc,\psi_{m},\;\xi\;\in \real[\vec X]$, and }
 \\
 &
 \text{maximum degrees $d_{1},\dotsc, d_{n}, e_{1},\dotsc, e_{m}\in \nat$}
  \\
   \text{\bfseries Output: \quad}&
  \text{SOSs 
 \begin{math}
     s_{1}\in \cone(\emptyset)_{\le d_{1}},
  \dotsc,
  s_{n}\in \cone(\emptyset)_{\le d_{n}}
 \end{math}
 and}
 \\
 &
 \text{polynomials 
 \begin{math}
    t_{1}\in \real[\vec X]_{\le e_{1}},
  \dotsc,
  t_{m}\in \real[\vec X]_{\le e_{m}}
 \end{math}
  }
 \\
 &\text{such that
 $s_1\varphi_1+\cdots + s_n \varphi_n + t_1\psi_1+ \cdots +t_m\psi_m +\xi= 0$
 }
 \end{tabular}
\end{displaymath}
Here $\real[\vec X]_{\le e}$ denotes the set of polynomials in $\vec{X}$ whose degree is no bigger than $e$. Similarly 
 $\cone(\emptyset)_{\le d}$ is the set of SOSs with degree $\le d$. 
\end{mydefinition}
The problem $\FindPoly$ is principally about finding polynomials
$s_{i}, t_{j}$
subject to
$\sum_{i}s_{i}\varphi_{i}
+\sum_{j}t_{j}\psi_{j}
+ \xi = 0$; this problem is known as \emph{polynomial Diophantine equations}. In $\FindPoly$   SOS requirements are additionally imposed on part of a solution (namely $s_{i}$);  degrees are bounded, too, for algorithmic purposes. 

In Algorithm~1 we rely on Thm.~\ref{thm:FromCertificateToInterpolant} to generate an interpolant: roughly speaking, one looks for a disjointness certificate
$f,f',g, h, h'$ within a predetermined maximum degree $b$. 
This search is relaxed to an instance of 
 $\FindPoly$ (Def.~\ref{def:findPoly}), with $n =2^{s}+2^{s'}$, $m=u+u'$, and $\xi = 1+g^{2}$, as in Line~\ref{line:algDaiPDioph}. 
The last relaxation, introduced in~\cite{DaiXZ13}, is derived  from the following representation of elements of the cone $\cone(\overrightarrow{f}, \overrightarrow{f'})$,
the multiplicative monoid $\mathcal{M}(\overrightarrow{g}, \overrightarrow{g'})$ and the ideal $\mathcal{I}(\overrightarrow{h}, \overrightarrow{h'})$, respectively.
 \begin{itemize}
  \item Each element $h$ of $\mathcal{I}(\overrightarrow{h}, \overrightarrow{h'})$ is of the form $h=\sum_{j=1}^u \beta_j h_j + \sum_{j'=1}^{u'}\beta'_{j'} h'_{j'}$, where $\beta_{j}, \beta_{j'}\in \real[\vec X]$. This is a standard fact in ring theory.
  \item Each element of $\mathcal{M}(\overrightarrow{g},
	\overrightarrow{g'})$ is given by the product of finitely many
	elements from $\overrightarrow{g}, \overrightarrow{g'}$ (here
	multiplicity matters). In Algorithm~\ref{alg:Dai} a polynomial
	$g$  is fixed to a ``big'' one. This is justified as follows: in case the constraint~(\ref{eq:Dai}) is satisfiable using a smaller polynomial $g'$ instead of $g$, by multiplying the whole equality~(\ref{eq:Dai}) by $1+ (g/g')^{2}$ we see that (\ref{eq:Dai}) is satisfiable using $g$, too. 
  \item For the cone $\cone(\overrightarrow{f}, \overrightarrow{f'})$ we use the following fact (here $\mathbf 2=\{0,1\}$). The lemma seems to be widely known but we present a proof in Appendix~\ref{subsec:ConeSOSRepr} for the record. 
  \end{itemize}
	\begin{mylemma}
  \label{lem:ConeSOSRepr}
  An arbitrary element $f$ of the cone $\cone(f_1,\dots,f_s)$ can be expressed  as
  $f=\sum_{i\in \mathbf{2}^s} \alpha_{i} f_1^{i_1}\dots f_s^{i_s}$, using
  SOSs $\alpha_i$ (where $i\in \mathbf{2}^s$).
 \myqed
	\end{mylemma}
The last representation justifies the definition of $f$ and $h$ in Algorithm~\ref{alg:Dai} (Line~\ref{line:algDaiDefFH}). We also observe that  Line~\ref{line:algDaiReturn} of Algorithm~\ref{alg:Dai} corresponds to~(\ref{eq:daiEtAlInterpolant}) of Thm.~\ref{thm:FromCertificateToInterpolant}. 

In implementing 
Algorithm~\ref{alg:Dai} the following fact is crucial
(see~\cite[\S{}3.5]{DaiXZ13} and
also~\cite{Parrilo00,Parrilo03} for details):
the problem $\FindPoly$ (Def.~\ref{def:findPoly}) can be reduced to an SDP problem, the latter allowing an efficient solution by state-of-the-art SDP solvers. It should be noted, however, that numerical errors (inevitable in interior point methods) can pose a serious issue for our application:
 the constraint~(\ref{eq:Dai}) is an equality and hence fragile.

\auxproof{
\begin{myremark}\label{rem:ParriloAndDai}
Overall, Algorithm~\ref{alg:Dai}  relaxes the problem of finding a certificate
 $f,f',g, h, h'$ in Thm.~\ref{thm:FromCertificateToInterpolant} into
 a suitable instance of $\FindPoly$, which is  translated to
 a suitable SDP problem, which is in turn solved by an SDP solver. This workflow  
 in~\cite{DaiXZ13} is derived from one in~\cite{Parrilo00} for infeasibility check of an \SASdiseq\ $\mathcal{T}$ (see also~\cite[Thm.~5.1]{Parrilo03}), where search for a Positivstellensatz certificate  ($f,g,h$ in Thm.~\ref{thm:positivStellenSatz})  undergoes similar  SDP relaxation. The latter method in~\cite{Parrilo00,Parrilo03} for infeasibility check turns out to be complete, while its variant for interpolation in~\cite{DaiXZ13} is not. 
Later in Rem.~\ref{rem:completenessOfOurStrictConeAlgorithm} we continue this discussion.
\end{myremark}
}

\section{Positivstellensatz and Interpolation, Revisited}\label{sec:limitationOfDai}

\subsection{Analysis of the Interpolation Algorithm by Dai et al.}\label{subsec:limitationOfDai}
Intrigued by its solid mathematical foundation in real algebraic
  geometry as well as its efficient implementation that exploits 
  state-of-the-art SDP solvers, we studied the framework by Dai et
  al.~\cite{DaiXZ13} (it was sketched
  in~\S{}\ref{subsec:DaiEtAlAlgorithm}). 
  In its course we obtained the following observations that motivate our current technical contributions.

 We first observed that  Algorithm~\ref{alg:Dai} from~\cite{DaiXZ13}
 fails to find ``sharp'' interpolants for ``barely disjoint'' predicates
 (see Example~\ref{ex:sharpInterpolantIntro}).
 This 
 turns out to be a general phenomenon (see Prop.~\ref{prop:kojima}).

 \begin{mydefinition}[symbolic closure]
  \label{def:symbolicclosure}
  Let $\mathcal T$ be the \SASdiseq's in~(\ref{eq:SAS}).
  The \emph{symbolic closure} 
  ${\mathcal{T}}_{\bullet}$
of $\mathcal T$ is 
  the \SASdiseq\ that is obtained by dropping all the disequality constraints
 $g_{j}(\vec{x})\neq 0$ 
 in $\mathcal{T}$.
  \begin{align}
    \label{eq:tildeT}
     \footnotesize
    \begin{array}{l}
       {\mathcal{T}}_{\bullet}
      =
      \left(
      \begin{array}{l}
        f_{1}(\vec{X},\vec Y) \ge 0\enspace,
       \;\dotsc,\;
       f_{s}(\vec{X},\vec Y) \ge 0\enspace,\;
       h_{1}(\vec{X},\vec Y) = 0\enspace,
       \;\dotsc,\;
       h_{u}(\vec{X},\vec Y) = 0
      \end{array}
      \right)
    \end{array}
    \end{align}
 \end{mydefinition}
 The intuition of symbolic closure of $\mathcal{T}$ is to replace all  strict inequalities $g'_{j}(\vec{X},\vec Y)>0$ in $\mathcal{T}$  with the corresponding non-strict ones  $g'_{j}(\vec{X},\vec Y)\ge 0$. Since only $\ge, \neq$ and $=$ are allowed in \SASdiseq's, strict inequalities $g'_{j}(\vec{X},\vec Y)>0$ are presented in the \SASdiseq\ $\mathcal{T}$ by using both
$g'_{j}(\vec{X},\vec Y)\ge 0$ and $g'_{j}(\vec{X},\vec Y)\neq 0$. The last definition drops the latter disequality ($\neq$) requirement. 

 The notion of symbolic closure most of the time coincides with  closure with respect to the usual Euclidean topology, but not in some singular cases. See Appendix~\ref{sec:closures}.


 \begin{mydefinition}[bare disjointness]
  \label{def:barelydisjoint}
  Let $\mathcal T$ and $\mathcal T'$ be \SASdiseq's.
  $\mathcal T$ and $\mathcal T'$ are \emph{barely disjoint}
  if $\sem{\mathcal T} \cap \sem{\mathcal T'} = \emptyset$ and
  $\sem{\mathcal T_\bullet} \cap \sem{\mathcal T'_\bullet} \neq \emptyset$.

  An interpolant $\mathcal{S}$ of barely disjoint \SASdiseq's  $\mathcal T$ and $\mathcal T'$  shall be said to be \emph{sharp}. 
 \end{mydefinition}

An example of barely disjoint \SASdiseq's\ is in Example~\ref{ex:sharpInterpolantIntro}:
 $(0,0)\in\sem{\mathcal{T}_{\bullet}}\cap \sem{\mathcal{T}'_{\bullet}}\neq
 \emptyset$.

 Algorithm~\ref{alg:Dai} does not work if the \SASdiseq's $\mathcal{T}$ and $\mathcal{T}'$ are only 
 barely disjoint.
In fact, such failure is theoretically guaranteed, 
as the following result states. 
Its proof (in Appendix~\ref{subsec:propKojima}) is much like for Thm.~\ref{thm:FromCertificateToInterpolant}. 
\begin{myproposition}
\label{prop:kojima}
 Let $\mathcal{T}$ and $\mathcal{T}'$ be the \SASdiseq's in~(\ref{eq:T}).
If  ${\mathcal{T}}$ and $\mathcal{T}'$ are barely disjoint  (in the sense of Def.~\ref{def:barelydisjoint}), 
there do not exist polynomials
 $\tilde f\in\cone(\overrightarrow{f},\overrightarrow{f'})$, 
$ g\in\multMon(\overrightarrow{g},\overrightarrow{g'})$ 
and
 $\tilde h\in \ideal(\overrightarrow{h},\overrightarrow{h'})$ 
such that $1+\tilde f+ g^2+\tilde h=0$.
\myqed
\end{myproposition}
\noindent
The conditions in Prop.~\ref{prop:kojima} on the polynomials $\tilde{f}, g, \tilde{h}$ are those for disjointness certificates for $\mathcal{T}$ and $\mathcal{T}'$ (Thm.~\ref{thm:FromCertificateToInterpolant}). As a consequence:  if $\mathcal{T}$ and $\mathcal{T}'$ are
only barely disjoint, 
interpolation relying on Thm.~\ref{thm:FromCertificateToInterpolant}---that underlies the framework in~\cite{DaiXZ13}---never succeeds. 


\subsection{Interpolation via Positivstellensatz, Sharpened}
\label{subsec:sharperInterpolationViaPositivstellensatz}
The last observation motivates our ``sharper'' variant of
Thm.~\ref{thm:FromCertificateToInterpolant}---a technical contribution that we shall present
shortly in Thm.~\ref{thm:scone}.
We switch  input formats by replacing disequalities $\neq$
(Def.~\ref{def:SAS}) with  $<$. This small change turns out to be useful when we formulate our
main result (Thm.~\ref{thm:scone}). 
\begin{mydefinition}[\SASstrineq]\label{def:SAS2}
A \emph{semialgebraic system with strict inequalities (\SASstrineq)} $\mathcal{T}$, in variables  $X_1,X_2,\dots,X_k$,  is
a sequence 
  \begin{equation}\label{eq:SAS2}\small
 \mathcal{T}
\;=\;
\left(
\begin{array}{l}
   f_{1}(\vec{X}) \ge 0\enspace,
  \;\dotsc,\;
  f_{s}(\vec{X}) \ge 0\enspace,
  \qquad
  g_{1}(\vec{X}) > 0\enspace,
  \;\dotsc,\;
  g_{t}(\vec{X}) > 0\enspace,
  \\
  h_{1}(\vec{X}) = 0\enspace,
  \;\dotsc,\;
  h_{u}(\vec{X}) = 0
\end{array}
\right)\enspace
\end{equation}
of 
 inequalities $f_{i}(\vec{X})\ge 0$, 
\emph{strict inequalities} $g_{j}(\vec{X})>  0$ and
equalities $h_{k}(\vec{X})=0$. 
Here $f_{i}, g_{j}, h_{k}\in \real[\vec{X}]$ are polynomials;
 $\sem{\mathcal{T}}\subseteq\real^{k}$ is defined like in Def.~\ref{def:SAS}. 
\end{mydefinition}


\SASstrineq's have the same expressive power as \SASdiseq's, as witnessed by the following  mutual translation. For the \SASdiseq\ $\mathcal{T}$ in~(\ref{eq:SAS}), the \SASstrineq\ 
\begin{math}
 \tilde{\mathcal{T}} \coloneqq
 \bigl(\, f_{i}(\vec{X}) \ge 0,
\,
  g^2_{j}(\vec{X}) > 0,
\,
h_{k}(\vec{X}) = 0
\,\bigr)_{i,j,k}
\end{math}
satisfies $\sem{\mathcal{T}}=\sem{\tilde{\mathcal{T}}}$. 
Conversely, for the 
\SASstrineq\ $\mathcal{T}$ in~(\ref{eq:SAS2}), the \SASdiseq\
\begin{math}
 \widehat{\mathcal{T}} \coloneqq
 \bigl(\, f_{i}(\vec{X}) \ge 0,
\,
  g^2_{j}(\vec{X}) \ge 0,
\,
  g^2_{j}(\vec{X}) \neq 0,
\,
h_{k}(\vec{X}) = 0
\,\bigr)_{i,j,k}
\end{math}
satisfies $\sem{\mathcal{T}}=\sem{\widehat{\mathcal{T}}}$.




One crucial piece for Positivstellensatz
was the closure properties of inequalities/disequalities/equalities encapsulated in the notions of cone/multiplicative monoid/ideal (Lem.~\ref{lem:closurePropertiesOfCMI}). We  devise a counterpart for strict inequalities.

\vspace{.5em}
\noindent
\begin{minipage}{\textwidth}
 \begin{mydefinition}[strict cone]
 \label{def:strictCone}
 A set $S \subseteq \real[\vec X]$ is a \emph{strict cone} if it
 satisfies the following closure properties:
 1) $f,g\in S$ implies $f+g \in S$;
 2) $f,g\in S$ implies $fg\in S$; and
 3) $r \in S$ for any positive real $r\in \real_{>0}$.
 For a subset $A$ of $\real[\vec{X}]$, we write $\scone(A)$ for the smallest strict cone that includes $A$.
 \end{mydefinition}
\end{minipage}

\vspace{.5em}
\noindent
\begin{minipage}{\textwidth}
\begin{mylemma}\label{lem:closurePropertiesOfS}
   Let $\vec x\in \real^{k}$ and $g_{j}\in \real[\vec X]$.
 If  $\vec{x} \in \sem{g_{1} > 0,\dots,g_{t} > 0}$, then 
 $g(\vec{x}) > 0$ for all $g \in \scone(g_1,\dots,g_t)$.
\myqed
\end{mylemma}
\end{minipage}

\vspace{.5em}

We can now formulate adaptation of Positivstellensatz. Its proof is in Appendix~\ref{subsec:ProofthmpositivStellenSatzStrict}.

\vspace{.5em}
\noindent
\begin{minipage}{\textwidth}
\begin{mytheorem}[Positivstellensatz for \SASstrineq]\label{thm:positivStellenSatzStrict}
 Let $\mathcal{T}$ be the \SASstrineq\ in~(\ref{eq:SAS2}).
 It is infeasible (i.e.\ $\sem{\mathcal{T}}=\emptyset$) if and only if there exist
 $f\in \cone(f_{1},\dotsc,f_{s},g_1, \dotsc, g_{t})$,
 $g\in \scone(g_{1},\dotsc,g_{t})$ and
 $h\in \ideal(h_{1},\dotsc,h_{u})$
 such that $f+g+h= 0$. 
 \qed
\end{mytheorem}
\end{minipage}


\vspace{.5em}
\noindent
From this we derive the following adaptation of Thm.~\ref{thm:FromCertificateToInterpolant} that allows to synthesize sharp interpolants. The  idea is as follows. In Thm.~\ref{thm:FromCertificateToInterpolant}, the constants $1$ (in~(\ref{eq:1plus})) and $1/2$ (in~(\ref{eq:daiEtAlInterpolant})) are there to enforce strict positivity. This is a useful trick but sometimes too ``dull'': one can get rid of these constants
and still make the proof of Thm.~\ref{thm:FromCertificateToInterpolant} work, for example when $g(\vec x)$ happens to belong to $\multMon(g_1,\dots,g_t)$
instead of $\multMon(g_1,\dots,g_t,g'_1,\dots,g'_{t'})$.

\begin{mytheorem}[disjointness certificate from strict cones]
\label{thm:scone}
 Let $\mathcal{T}$ and $\mathcal{T}'$ be the following \SASstrineq's, where $\vec X$ denotes the variables that occur in both of  $\mathcal{T},\mathcal{T}'$. 
\begin{align}
  \label{eq:Tscone}
  \footnotesize
  \begin{split}
    \mathcal{T}
  &=
  \left(
  \begin{array}{l}
    f_{1}(\vec{X},\vec Y) \ge 0\enspace,
   \;\dotsc,\;
   f_{s}(\vec{X},\vec Y) \ge 0\enspace,
   \quad
   g_{1}(\vec{X},\vec Y) > 0\enspace,
   \;\dotsc,\;
   g_{t}(\vec{X},\vec Y) > 0\enspace,
   \\
   h_{1}(\vec{X},\vec Y) = 0\enspace,
   \;\dotsc,\;
   h_{u}(\vec{X},\vec Y) = 0
  \end{array}
  \right),\\
  \mathcal{T}'
  &=
  \left(
  \begin{array}{l}
    f'_{1}(\vec{X},\vec Z) \ge 0\enspace,
   \;\dotsc,\;
   f'_{s'}(\vec{X},\vec Z) \ge 0\enspace,
   \quad
   g'_{1}(\vec{X},\vec Z) > 0\enspace,
   \;\dotsc,\;
   g'_{t'}(\vec{X},\vec Z) > 0\enspace,
   \\
   h'_{1}(\vec{X},\vec Z) = 0\enspace,
   \;\dotsc,\;
   h'_{u'}(\vec{X},\vec Z) = 0
  \end{array}
  \right).
  \end{split}
\end{align}

Assume there exist
 \begin{align}
&\begin{array}{l}
f\in \cone(f_1,\dots,f_s,g_1,\dots,g_t)\enspace, \quad
f'\in\cone(f_1',\dots,f_{s'}',g'_1,\dots,g'_{t'})\enspace,
\\
g\in\scone(g_1,\dots,g_t)\enspace, \quad
h\in \ideal(h_1,\dots,h_u)\enspace,\quad\text{and}\quad
h'\in
\ideal(h'_1,\dots,h'_{u'})
\end{array} 
\notag
\\
&\quad\text{such that}\quad
 f+f'+g+h+h'\;=\;0\enspace.
\label{eq:thmsconeConstraint}
\end{align}
Then the \SASstrineq's $\mathcal{T}$ and $\mathcal{T}'$ are disjoint. Moreover the \SASstrineq
\begin{align}\label{eq:thmsconeInterpolant}
\mathcal{S}\;\coloneqq\;(f+g+h>0)
\end{align}
satisfies the conditions of an interpolant of $\mathcal{T}$ and $\mathcal{T}'$ (Def.~\ref{def:interpolant}), except for Cond.~3.\ (the common variable condition). 
\myqed
\end{mytheorem}
%

The proof   is  like for Thm.~\ref{thm:FromCertificateToInterpolant}. 
We also have the following symmetric variant.

\vspace{.5em}

\noindent
\begin{minipage}{\textwidth}
\begin{mytheorem}\label{thm:scone2}
Assume the conditions of Thm.~\ref{thm:scone}, but let us now require
 $g\in \scone(g'_1,\dots,g'_{t'})$ (instead of $g\in \scone(g_1,\dots,g_t)$). Then $\mathcal{S}=(f+h\ge 0)$ is an interpolant of $\mathcal{T}$ and $\mathcal{T'}$ (except for the common variable condition). 
\myqed
\end{mytheorem}
\end{minipage}

\vspace{.5em}
\noindent
\begin{minipage}{\textwidth}
\begin{myexample}
 Let us apply Thm.~\ref{thm:scone} to
 \begin{math}
  \mathcal{T} = (-y>0)
 \end{math}
 and
 \begin{math}
 \mathcal{T}' = (y-x\ge 0,y+x\ge 0)
 \end{math} (these are only barely disjoint).
 There exists a disjointness certificate $f,f',g,h,h'$: indeed, we can take 
 $f=0\in \cone(-y)$,
 $f'=2y = (y - x) + (y + x)\in \cone(y-x,y+x)$, $g=2(-y)
 \in \scone(-y)$, and $h=h'=0\in \ideal(\emptyset)$; for these we have 
 $f+f'+g+h+h'=0$. This way an interpolant $\mathcal{S}=(f+g + h > 0)=(-2y > 0)$ is derived. 
\end{myexample}
\end{minipage}
 
\vspace{.5em}
\noindent
\begin{minipage}{\textwidth}
 \begin{myremark}\label{rem:thmscone}
Our use of strict cones allows to use a polynomial $g$ in~(\ref{eq:thmsconeInterpolant}).
This is in contrast with $g^{2}$ in~(\ref{eq:daiEtAlInterpolant}) and yields an interpolant of a potentially smaller degree.

 \end{myremark}
\end{minipage}

\vspace{.5em}

\begin{algorithm}[tbp]
  \caption{Our interpolation algorithm  based on Thm.~\ref{thm:scone}.  Here 
  $\mathbf 2=\{0,1\}$ and $\sigma(b)=\{(k_1,\dots,k_t)\in \nat^t\mid k_1+\dots+k_t \le b+1\}$}
  \label{alg:scone}
  \begin{algorithmic}[1]
    \STATE \textbf{input}:
   \SASstrineq's $\mathcal{T},\mathcal{T}'$ in~(\ref{eq:Tscone}), and $b\in \nat$ (the maximum degree)
    \STATE \textbf{output}: 
   either an interpolant $\mathcal{S}$ of $\mathcal{T}$ and $\mathcal{T}'$, or $\NULL$
    \STATE   \label{line:scone2ReductionToFindPoly}
   Solve (an extension of) $\FindPoly$ to find 
   $(
   \overrightarrow{\alpha},
   \overrightarrow{\alpha'},
   \overrightarrow{\beta},
   \overrightarrow{\beta'},
      \overrightarrow{\gamma}
)$. Here:
\begin{itemize}
 \item  $\alpha_{ij} \in \cone(\emptyset)_{\le b}$ (for $i\in \mathbf{2}^s$, $j\in\mathbf{2}^{t}$) and
     $\alpha'_{i',j'} \in \cone(\emptyset)_{\le b}$ (for $i'\in \mathbf{2}^{s'}$, $j'\in\mathbf{2}^{t'}$) are SOSs, 
 \item  $\beta_{j}\in \real[\vec X]_{\le b}$ (for $j\in [1,u]$) and
     $\beta'_{j'}\in \real[\vec X]_{\le b}$ (for $j'\in [1,u']$) are polynomials, 
 \item and $\gamma_{k}\in \real_{\ge 0}$ (for $k\in \sigma(b)$) are nonnegative real numbers,
\end{itemize}
that are subject to the constraints
\begin{align}
&\begin{array}{l}
    \sum_{i\in \mathbf{2}^{s},\, j\in \mathbf{2}^{t}} \alpha_{ij} f_1^{i_1}\dotsm f_s^{i_s} g_1^{j_1}\dotsm g_t^{j_t}\\
+\;
\sum_{i'\in \mathbf{2}^{s'},\, j'\in \mathbf{2}^{t'}} \alpha'_{i'j'} {f'_1}^{i'_1}\dotsm {f'}_{s'}^{i'_{s'}} {g'}_1^{j'_1}\dotsm {g'}_{t'}^{j'_{t'}}
\\
 +\;
\sum_{k\in\sigma(b)}
\gamma_k g_1^{k_1}\dotsm g_t^{k_t}
\;+\;
\sum_{j=1}^u \beta_j h_j
\;+\;
 \sum_{j'=1}^{u'}\beta'_{j'} h'_{j'} 
\quad=\quad 0\enspace,
\end{array} 
\label{eq:eqconstraintAlgo2_1} 
\\
&\textstyle \sum_{k\in\sigma(b)}\gamma_k \ge 1\enspace, \quad\text{and}
\label{eq:eqconstraintAlgo2_2} 
\\
& \text{some equality constraints that forces the common variable condition.
}
\label{eq:eqconstraintAlgo2_3} 
\end{align}
(Such 
   $(
   \overrightarrow{\alpha},
   \overrightarrow{\alpha'},
   \overrightarrow{\beta},
   \overrightarrow{\beta'},
      \overrightarrow{\gamma}
)$ may not be found, 
in which case return $\NULL$)
   \STATE \label{line:scone2PrepareFGH}
    $f\coloneqq
    \sum_{i\in \mathbf{2}^{s},\, j\in \mathbf{2}^{t}} \alpha_{ij} f_1^{i_1}\dotsm f_s^{i_s} g_1^{j_1}\dotsm g_t^{j_t}
$;
   $g\coloneqq \sum_{k\in \sigma(b)}
\gamma_k g_1^{k_1}\dotsm g_t^{k_t}
$;
   $h\coloneqq \sum_{j=1}^u \beta_j h_j$

   \RETURN    \label{line:scone2Return}
   $\mathcal{S} \coloneqq \left(f +g + h > 0\right)$

  \end{algorithmic}
\end{algorithm}

We derive an interpolation algorithm from Thm.~\ref{thm:scone}; see~Algorithm~\ref{alg:scone}. An algorithm based on Thm.~\ref{thm:scone2} can be derived similarly, too.

Algorithm~\ref{alg:scone} reduces search for a disjointness certificate $f,f',g,h,h'$ (from Thm.~\ref{thm:scone}) to a problem similar to $\FindPoly$ (Line~\ref{line:scone2ReductionToFindPoly}). Unlike the original definition of $\FindPoly$ (Def.~\ref{def:findPoly}), here we impose additional constraints 
(\ref{eq:eqconstraintAlgo2_2}--\ref{eq:eqconstraintAlgo2_3}) other than the equality (\ref{eq:eqconstraintAlgo2_1}) that comes from~(\ref{eq:thmsconeConstraint}). It turns out that, much like $\FindPoly$ allows relaxation to SDP problems~\cite{DaiXZ13,Parrilo00,Parrilo03}, the problem  in 
Line~\ref{line:scone2ReductionToFindPoly} can also be reduced to an SDP problem. 

 The constraint
(\ref{eq:eqconstraintAlgo2_2}) is there to force $g=\sum_{k\in \mathbf{b}^{t}}
\gamma_k g_1^{k_1}\dotsm g_t^{k_t}
$
(see~(\ref{line:scone2PrepareFGH}))
 to belong to the \emph{strict} cone $\scone(g_1,\dots,g_t)$.  A natural requirement $\sum_{k\in \mathbf{b}^{t}}\gamma_k > 0$ for that purpose does not allow encoding to an SDP constraint so we do not use it. Our relaxation from $\sum_{k\in \mathbf{b}^{t}}\gamma_k>0$ to $\sum_{k\in \mathbf{b}^{t}}\gamma_k\ge 1$ is inspired by~\cite{Rybalchenko2007}; it does not lead to loss of generality in our current task of finding polynomial certificates.

The constraints~(\ref{eq:eqconstraintAlgo2_3}) are extracted in the following straightforward manner: we look at the coefficient of each monomial in $f+g+h$ (see Line~\ref{line:scone2Return}); and for each monomial that involves variables other than $\vec X$ we require the coefficient to be equal to $0$. The constraint is linear in the SDP variables, that we can roughly consider as the coefficients of the monomials in $   \overrightarrow{\alpha},
   \overrightarrow{\alpha'},
   \overrightarrow{\beta},
   \overrightarrow{\beta'},
      \overrightarrow{\gamma}
$. 

 Derivation of Algorithm~\ref{alg:scone} from Thm.~\ref{thm:scone2} also relies on the following analogue of Lem.~\ref{lem:ConeSOSRepr}. Its proof is 
in Appendix~\ref{subsec:SConeSOSRepr}.
\begin{mylemma}
  \label{lem:SConeSOSRepr}
  An arbitrary element of the strict cone $\scone(f_1,\dots,f_s)$ can be expressed as
  $\sum_{i\in \nat^s} \alpha_{i} f_1^{i_1}\dotsm f_s^{i_s}$, where
   $\alpha_i \in \real_{\ge 0}$ are  nonnegative reals
 (for $i\in \nat^s$)
  such that:  there exists $i$ such that $\alpha_{i}>0$; and  $\alpha_i\neq 0$ for only finitely many $i$. 
 \myqed
\end{mylemma}



To summarize: our analysis of the framework of~\cite{DaiXZ13} has led to a new algorithm (Algorithm~\ref{alg:scone}) that allows ``sharp'' interpolation of barely disjoint SASs. This algorithm is based on strict inequalities ($>$) instead of disequalities ($\neq$); we introduced the corresponding notion of \emph{strict cone}. The algorithm allows solution by numeric SDP solvers. Moreover we observe that the common variable condition---that  seems to be only partially addressed in~\cite{DaiXZ13}---allows encoding as SDP constraints.

We conclude by noting that our algorithm (Algorithm~\ref{alg:scone}) generalizes Algorithm~\ref{alg:Dai} from~\cite{DaiXZ13}. More specifically, given \SASdiseq's $\mathcal{T}$ and $\mathcal{T}'$ that are disjoint, if Algorithm~\ref{alg:Dai} finds an interpolant, then Algorithm~\ref{alg:scone} also finds an interpolant after suitable translation of 
$\mathcal{T}$ and $\mathcal{T}'$ to \SASstrineq's. See Appendix~\ref{sec:generalization}.


\auxproof{
\begin{myremark}\label{rem:completenessOfOurStrictConeAlgorithm}
Algorithm~\ref{alg:scone} (based on Thm.~\ref{thm:scone}) is sharper than
Algorithm~\ref{alg:Dai} from~\cite{DaiXZ13}, in the sense that we can now interpolate SASs like in Example~\ref{ex:sharpInterpolantIntro}. This does not mean that we recover completeness that is lost in the transition from infeasibility check~\cite{Parrilo00,Parrilo03} to interpolation~\cite{DaiXZ13} (see Rem.~\ref{rem:ParriloAndDai}). 

Specifically, assuming that $\mathcal{T}$ and 
 $\mathcal{T}'$  are disjoint, 
Thm.~\ref{thm:positivStellenSatzStrict} always yields an infeasibility certificate $\tilde f\in \cone(\overrightarrow{f},\overrightarrow{f'},\overrightarrow{g},\overrightarrow{g'})$, $\tilde g\in \scone(\overrightarrow{g},\overrightarrow{g'})$ and $\tilde h\in \mathcal{I}(\overrightarrow{h},\overrightarrow{h'})$. In order to obtain an interpolant like in Thm.~\ref{thm:scone}, however, we have to be able to ``split'' $\tilde f, \tilde g, \tilde h$ into  parts that can be attributed to $\mathcal{T}$ (and to $\mathcal{T}'$) only. This additional requirement is a main source of incompleteness of Thm.~\ref{thm:scone} (and of Thm.~\ref{thm:FromCertificateToInterpolant} as well).
\end{myremark}
}

\section{Implementation: Numerical Errors and Rounding}\label{sec:impl}
Our implementation, that we named \SSPolyInt\ (\emph{Sharp and Simple Interpolants}), is essentially Algorithm~\ref{alg:scone}; in it we use an SDP solver to solve Line~\ref{line:scone2ReductionToFindPoly}. Specifically we use the SDP solver SDPT3~\cite{Toh99} via YALMIP as the backend. 

The biggest issue in the course of implementation is \emph{numerical errors}---they are inevitable due to (numerical) interior point methods that underlie most state-of-the-art SDP solvers. For one thing, we often get incorrect interpolants due to numerical errors (Example~\ref{ex:invalidCandidateIntro}). For another, in the context of  program verification  simpler interpolants are often more useful, reflecting simplicity of human insights (see~\S{}\ref{sec:intro}). Numerical solutions, on the contrary, do not very often provide humans with clear-cut understanding.

In our implementation we cope with numerical errors by \emph{rounding} numbers. More specifically we round \emph{ratios} $x_{1}: x_{2}: \dots : x_{n}$ because our goal is to simplify  a polynomial inequality $f+g+h>0$ (imagine the ratio between coefficients). For this purpose we employ an extension of \emph{continued fraction expansion}, a procedure used e.g.\ for the purpose of 
 \emph{Diophantine approximation}~\cite{lang1995Dioph} (i.e.\ finding the best rational approximation
$k_{1}/ k_{2}$  of a given real $r$). 
Concretely, our extension takes a ratio
 $x_{1}: \dots : x_{n}$ of natural numbers
 (and a parameter $d$ that we call \emph{depth}), and returns a simplified ratio 
 $y_{1}: \dots : y_{n}$.

 \begin{figure}[tbp]\textlb
 \begin{math}
  \xymatrix@C+2.5em@R=1em{
   {}
     \ar[r]^-{\mathcal{T},\mathcal{T'},b}
  &
   *+[F]{\txt{Alg.~2\\ (SDP)}}
     \ar[r]^-{\txt{original\\ candidate\\ $\vec v$}}_-{d\coloneqq 1}
  &
     *+[F]{\txt{rounding}}
     \ar[r]^-{\txt{rounded\\ candidate\\ $\vec v_{d}$}}
  &
     *+[F]{\txt{validation}}
     \ar[r]^-{\txt{pass}}
     \ar[d]^-{\txt{fail}}
  &
    {\text{interpolant}}
  \\
  &&&
     *+[F]{\vec{v}_{d}=\vec{v}?}
     \ar[r]^-{\txt{yes}}
     \ar`l[lu]^-{d:=d+1}_-{\txt{no}} [lu]
 &
    {\NULL}
 }
 \end{math} 
 \caption{The workflow of our tool \SSPolyInt}
 \label{fig:overviewOfThetool}
 \end{figure}
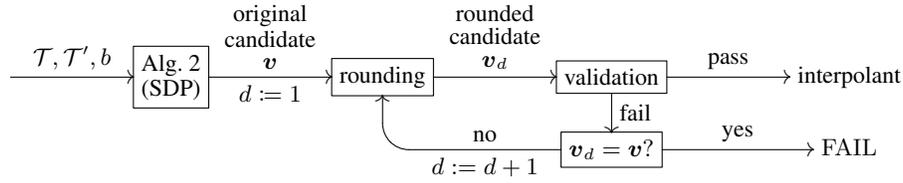

Overall the workflow of our tool \SSPolyInt\ is as in Fig.~\ref{fig:overviewOfThetool}. 
\begin{itemize}
 \item 
 We first run Algorithm~\ref{alg:scone}. Its output---more precisely the solution of the SDP problem in Line~\ref{line:scone2ReductionToFindPoly}---may not yield an interpolant, due to numerical errors. The output is therefore called a \emph{candidate} $\vec{v}$. 
 \item 
 We then round the candidate $\vec{v}$ iteratively, starting with the depth $d=1$ (the coarsest approximation that yields the simplest candidate $\vec{v}_{1}$) and incrementing the depth $d$. 
The bigger the depth $d$ is, the less simple and the closer to the original $\vec{v}$ the rounded candidate $\vec{v}_{d}$ becomes. 
 \item 
 In each iteration we check if the candidate $\vec{v}_{d}$  yields a valid interpolant. This \emph{validation} phase is conducted purely symbolically, ensuring soundness of our tool. 
 \item 
 Our rounding algorithm eventually converges and we have $\vec{v}_{d}=\vec{v}$ for a sufficiently large $d$ (Lem.~\ref{lem:convergenceOfExtDiophantineApprox}). In case we do not succeed by then we return $\NULL$. 
\end{itemize}
In other words, we try candidates $\vec{v}_{1},\vec{v}_{2},\dotsc$, from simpler to more complex, until we obtain a symbolically certified interpolant (or fail). This is the \emph{rounding and validation} workflow that we adopted from~\cite{Harrison2007}.

Our workflow involves another parameter $c\in \nat$ that we call \emph{precision}. It is used in an empirical implementation trick that we apply to the original candidate $\vec{v}$: we round it off to $c$ decimal places.

The tool \SSPolyInt\ is implemented in OCaml. When run with \SASstrineq's $\mathcal{T}$, $\mathcal{T'}$ (and a parameter $b\in\nat$ for the maximum degree, see Algorithm~\ref{alg:scone}) as input, the tool generates MATLAB code that conducts the workflow in Fig.~\ref{fig:overviewOfThetool}. The latter relies on the SDP solver SDPT3~\cite{Toh99} via YALMIP as the backend.


\subsection{Rounding}
\label{subsec:rounding}



\begin{algorithm}[tbp]
  \caption{Extended continued fraction expansion $\CFE$}
  \label{alg:approximate}
  \begin{algorithmic}[1]
    \STATE \textbf{input}: $x=(x_1,\dots,x_n) \in \nat^{n}$ (at least one  of $x_1,\dots,x_n$ is nonzero), and \emph{depth} $d\in \nat_{>0}$
   \STATE Pick $p$ so that $x_p$ is the smallest among the nonzero elements in $x_1,\dots,x_n$ 

         (say the smallest among such $p$'s)
    \STATE $a \coloneqq (\lfloor x_1/x_p \rfloor, \dots, \lfloor x_n/x_p\rfloor )$
    \IF{$d = 1$}
     \STATE $y \coloneqq a/\gcd(a)$
     \RETURN $y$
    \ELSE
   \STATE $r \coloneqq (x_1-a_1x_p,\,\dots,\,\banme{p-1}{x_{p-1}-a_{p-1}x_p},\,\banme{p}{x_p},\,\banme{p+1}{x_{p+1}-a_{p+1}x_p},\,\dots,\,x_n-a_nx_p) $
   \STATE $r'\coloneqq \CFE(r,d-1)$ 
   \qquad\qquad
   \COMMENT{a recursive call}
   \STATE $y \coloneqq (a_1 r'_p + r'_1,\dots,\banme{p-1}{a_{p-1}r'_p + r'_{p-1}},\banme{p}{r'_p},\banme{p+1}{a_{p+1}r'_p + r'_{p+1}},\dots,a_nr'_p + r'_n)$
   \RETURN $y/\gcd(y)$
    \ENDIF
  \end{algorithmic}
\end{algorithm}


\begin{wrapfigure}[4]{r}{2.5cm}
\vspace{-1.3cm}
 \scalebox{.65}{\begin{tabular}{c|c}
  $d$& $\CFE(x,d)$ \\
 \hline\hline 
  $1$& $(15,1,6)$ \\
  $2$& $(31,2,13)$ \\
  $3$& $(172,11,71)$ \\
  $4$& $(204,13,84)$ \\
  $5$& $(11515,735,4747)$ \\
  $6$& $(81389,5195,33552)$ \\
  $7$& $(174293,11125,71851)$ \\
  $8$& $(174293,11125,71851)$ \\
 \hline
  $x$& $(871465,55625,359255)$ 
 \end{tabular}
 }
 \end{wrapfigure}
\noindent
 \emph{Continued fraction expansion} is a well-known method for rounding a real number to a rational; it is known to satisfy an optimality condition called \emph{Diophantine approximation}. One can think of it as a procedure that simplifies ratios $x_{1}:x_{2}$ of two numbers.

In our tool we use our extension of the procedure that simplifies ratios  $x_{1}: \dots : x_{n}$. It is the algorithm $\CFE$ in Algorithm~\ref{alg:approximate}. 
An example is  in the above table, where
 $x=(871465,55625,359255)$.  
One sees that the ratio gets more complicated as the depth $d$ becomes bigger. For the depth  $d=7,8$ the output is equivalent to the input $x$.

Our 
 algorithm $\CFE$ enjoys the following pleasant
 properties. 
Their proofs are in Appendix~\ref{subsec:proofCFE}.

\vspace{.5em}
\noindent
\begin{minipage}{\textwidth}
\begin{mylemma}\label{lem:convergenceOfExtDiophantineApprox}
\begin{enumerate}
 \item (Convergence)
 The output  $\CFE(x,d)$   stabilizes for sufficiently large $d$; moreover the limit coincides with the input ratio $x$.  That is: for each $x$ there exists $M$ such that $\CFE(x,M)=\CFE(x,M+1)=\dots=x$ (as ratios).
 \item (Well-definedness) $\CFE$ respects  equivalence of ratios. That is, if $x,x'\in\nat^{n}$ represent the same ratio, then $\CFE(x,d)=\CFE(x',d)$ (as ratios) for each $d$.
 \myqed
\end{enumerate} 
\end{mylemma}
\end{minipage}
\vspace{.5em}

The algorithm $\CFE$ takes a positive ratio $x$ as input.  In the workflow in Fig.~\ref{fig:overviewOfThetool} $\CFE$ is applied to ratios with both positive and negative numbers; we deal with such input by first taking absolute values  and later adjusting signs. 



\subsection{Validation}\label{subsec:validation}
Potential unsoundness of verification  methods due to
numerical errors has been identified as a major challenge  (see e.g.~\cite{RouxVS16,Harrison2007,Besson2007,PEYRL2008269,KaltofenLYZ08,PlatzerQR09}). In our tool we enforce soundness (i.e.\ that the output is indeed an interpolant) by the validation phase in Fig.~\ref{fig:overviewOfThetool}. 

There the candidate $\vec{v}_{d}$ in question is fed back to the
constraints in (the SDP problem that is solved in)
Algorithm~\ref{alg:scone},\footnote{
In Algorithm~\ref{alg:scone} we introduced the constraint $\sum_{k\in \mathbf{b}^{t}}\gamma_k
\ge 1$ in~(\ref{eq:eqconstraintAlgo2_2})
 as  a relaxation of 
a natural constraint $\sum_{k\in \mathbf{b}^{t}}\gamma_k > 0$; see~\S{}\ref{subsec:sharperInterpolationViaPositivstellensatz}. 
In the validation phase of
our implementation we wind back the relaxation
$\sum_{k\in \mathbf{b}^{t}}\gamma_k
\ge 1$  to the original 
 constraint with $>0$.} and we check the constraints are satisfied. 
 The check must be symbolic. For equality constraints such symbolic check is easy. For semidefiniteness constraints, we rely on the following well-known fact: a symmetric real matrix $M$ is positive semidefinite if and only if all the principal minors of $M$ are nonnegative.  This characterization allows us to check semidefiniteness using only addition and multiplication.
 We find no computation in our validation phase to be overly expensive. This is in contrast with QE-based validation methods employed e.g.\ in~\cite{DaiGXZ17}: while symbolic and exact, the CAD algorithm for QE is known to be limited in scalability.



\section{Experiments}
\label{sec:experiments}
 We now present some experiment results. 
In the first part
we present some simple geometric examples that call for ``sharp'' interpolants; 
in the second
we discuss some program verification scenarios.  These examples  demonstrate our tool's capability of producing simple and sharp interpolants, together the benefits of such interpolants in program verification techniques.

The experiments were done on Apple MacBook Pro with
2.7 GHz Intel Core i5 CPU and 16 GB memory.
As we described in~\S{}\ref{sec:impl}, our tool \SSPolyInt\ consists of OCaml code
that generates MATLAB code; the latter runs the workflow in Fig.~\ref{fig:overviewOfThetool}.
Running the OCaml code
finishes in
milliseconds; running the resulting MATLAB code takes longer, typically for seconds. 
 The execution time shown here is
the average of 10 runs.

 Our tool has two parameters: the maximum degree $b$ and precision $c$ (\S{}\ref{sec:impl}). 
 In all our examples the common variable condition (in Def.~\ref{def:interpolant}) is successfully enforced. 

\vspace{.5em}
\noindent
\textbf{Geometric Examples}\quad
Table~\ref{tab:experimentResults} summarizes the performance of our
tool on interpolation problems.
For the input 6, we tried  parameters
$(b,c)=
(1,1),
(1,2),\dotsc,
(1,5)$
and $(2,5)$
but all failed, leading to $\NULL$ in Fig.~\ref{fig:overviewOfThetool}.
The input 9 contains disjunction, which is not allowed in \SASstrineq's.
It is dealt with using the technique described in~\cite[\S{}3.1]{DaiXZ13}:
an interpolant of 
$\mathcal{T}$ and $\mathcal{T}'$ is given by
$\bigvee_{i}\bigwedge_{j}\mathcal{S}_{ij}$, where $\mathcal{S}_{ij}$ is
an interpolant of each pair of disjuncts $\mathcal{T}_{i}$ and
$\mathcal{T}'_j$ of $\mathcal{T}$ and $\mathcal{T}'$, respectively.


\begin{table}[p]
  \centering
  \caption{Experiment results.  $\mathcal{T}$ and $\mathcal{T}'$ are
    inputs, and $\mathcal{S}$ is our output (see Fig.~\ref{fig:experimentResults} too).  The ``time'' column shows the
    execution time (in seconds) of the generated MATLAB code, $b$ and $c$ show the successful choice of parameters, and $d$ is the depth for which the workflow in
    Fig.~\ref{fig:overviewOfThetool} terminated.}
\scalebox{.85}{  \begin{tabular}{c||c|c||c||c|c|c|c}
    & $\mathcal{T}$ & $\mathcal{T}'$ & $\mathcal{S}$ & time [s] &
    $b$ & $c$ & $d$ \\\hline
    1 & $y>x, x>-y$ & $0 \ge y$ & $4y > 0$ & 2.19 & 0 & 5 & 1\\\hline
    2 & $y \le 0$ & $y > x^2$ & $-2y \ge 0$ & 5.68 & 2 & 3 & 1 \\\hline
    3 & $y > x, x > -y$ & $y \le x, x \le -y$ & $4y > 0$ &
    2.67 & 0 & 5 & 1 \\\hline
    4 & $y > x, x > -y$ & $y \le -x^2$ & $8y + 4x^2 > 0$ &
    5.09 & 2 & 1 & 1 \\\hline
    5 & $y \le -1$ & $x^2 + y^2 < 1$ & $34y^2 - 68y - 102 \ge 0$ &
    7.58 & 2 & 5 & 3 \\\hline
    6 & $x^2 + (y-1)^2 \le 1$ & $x^2 + (y-2)^2 > 4$ &
    FAIL & 14.0 & 2 & 5 & 8 \\\hline
    7 & $x^2 + (y+1)^2 \le 1$ & $x^2 + (y-1)^2 < 1$ &
    $\begin{array}{l}
       18x^2y - 14x^2y^2 - 144y\\
       + 28y^2 - 7x^4 + 18y^3 - 7y^4 \ge 0
     \end{array} $ &
    6.45 & 2 & 2 & 2 \\\hline
    8 & $x \ge z^2$ & $x < -y^2$ & $2x \ge 0$ &
    7.67 & 2 & 3 & 1 \\\hline
    9 & $
    \begin{array}{l}
      (y  \ge (x-1)^2) \lor {} \\ (y > (x+1)^2)
    \end{array}
    $ & 
    $ \begin{array}{l}
        (y  < -(x-1)^2) \lor {} \\ (y \le -(x+1)^2)
      \end{array}$ &
    \scalebox{0.8}{$\begin{array}{l}
       \left((586x + 293y + 119 > 0) \land
        (333y \ge 0)\right) \lor {} \\
        \left((333y > 0) \land
        (374y - 748x - 117 \ge 0)\right)
      \end{array}$} &
    43.7 & 2 & 3 & 3 
  \end{tabular}
}  \label{tab:experimentResults}
\end{table}

\begin{figure}[p]
\centering
\begin{minipage}{.19\textwidth}\centering
 \includegraphics[width=\textwidth]{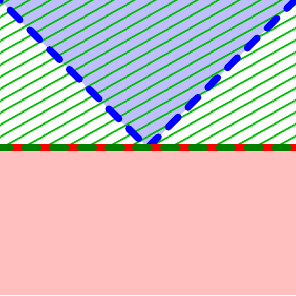}
 Input~1
\end{minipage} 
\begin{minipage}{.19\textwidth}\centering
 \includegraphics[width=\textwidth]{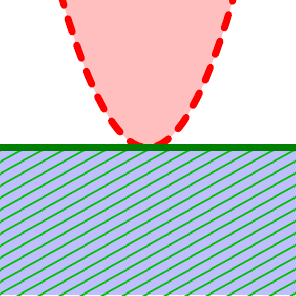}
 Input~2
\end{minipage} 
\begin{minipage}{.19\textwidth}\centering
 \includegraphics[width=\textwidth]{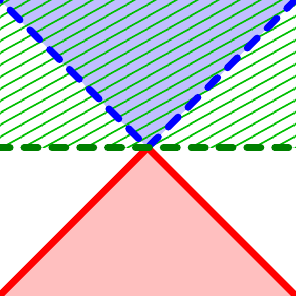}
 Input~3
\end{minipage} 
\begin{minipage}{.19\textwidth}\centering
 \includegraphics[width=\textwidth]{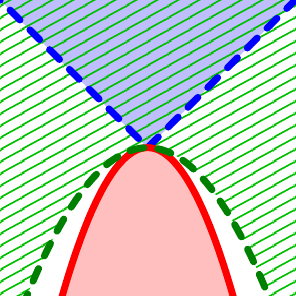}
 Input~4
\end{minipage} 
\begin{minipage}{.19\textwidth}\centering
 \includegraphics[width=\textwidth]{ex5.pdf}
 Input~5
\end{minipage} 
\begin{minipage}{.19\textwidth}\centering
 \includegraphics[width=\textwidth]{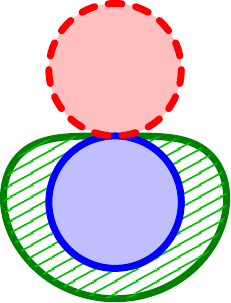}
 Input~7
\end{minipage} 
\begin{minipage}{.19\textwidth}\centering
 \includegraphics[width=\textwidth]{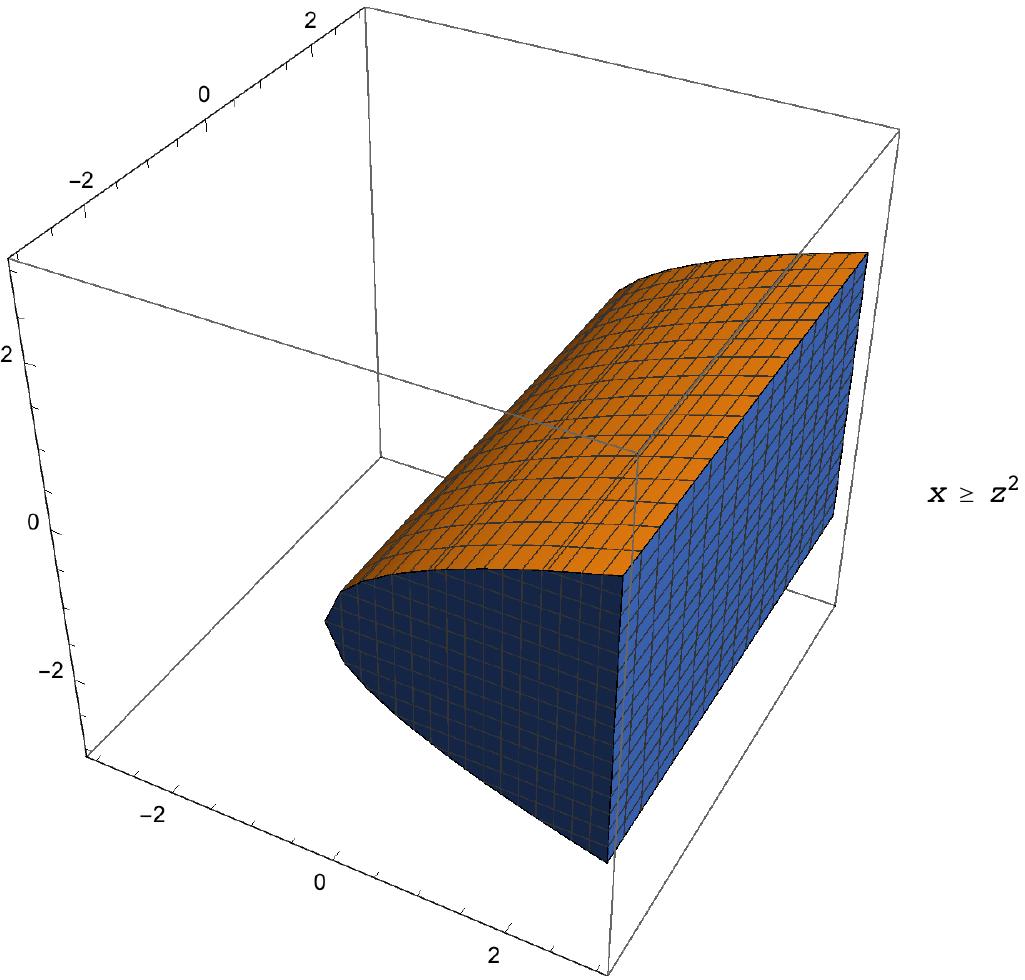}
 Input~8, $\cal T$
\end{minipage} 
\begin{minipage}{.19\textwidth}\centering
 \includegraphics[width=\textwidth]{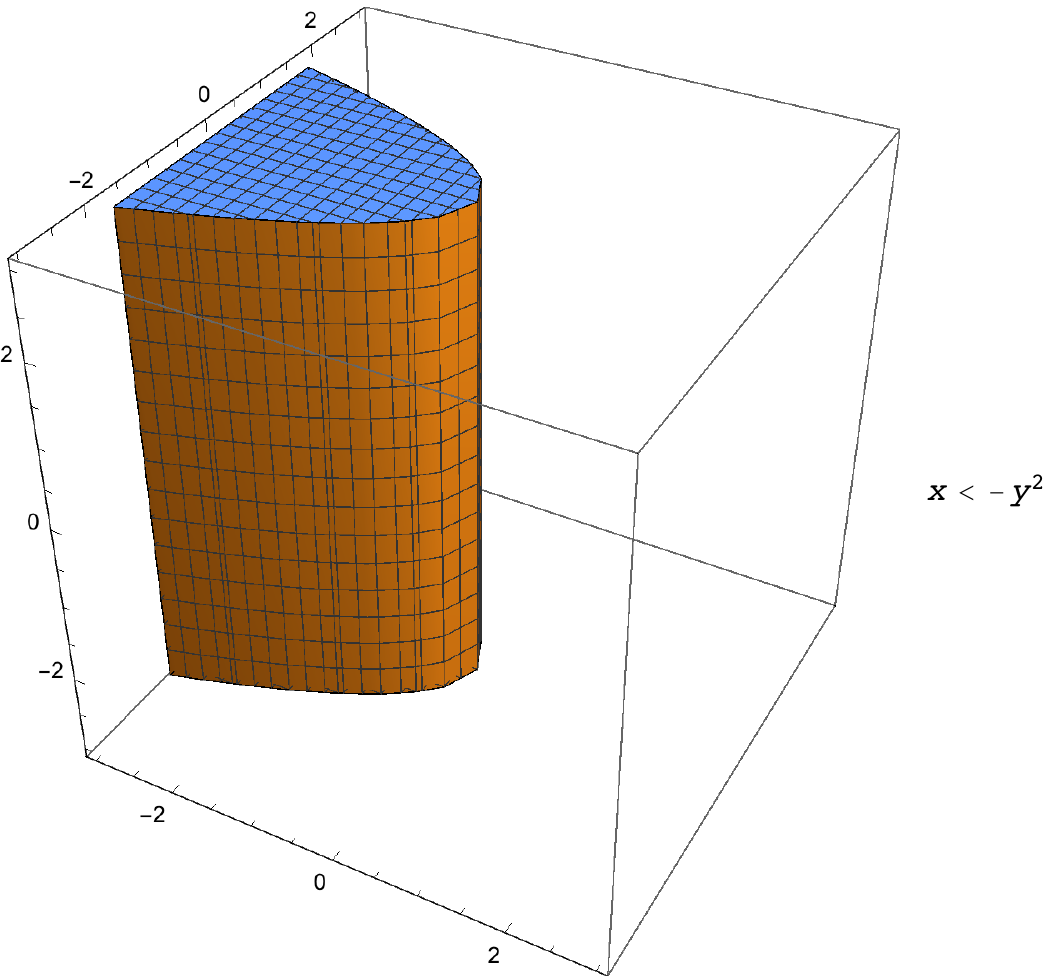}
 Input~8, ${\cal T}'$
\end{minipage} 
\begin{minipage}{.19\textwidth}\centering
 \includegraphics[width=\textwidth]{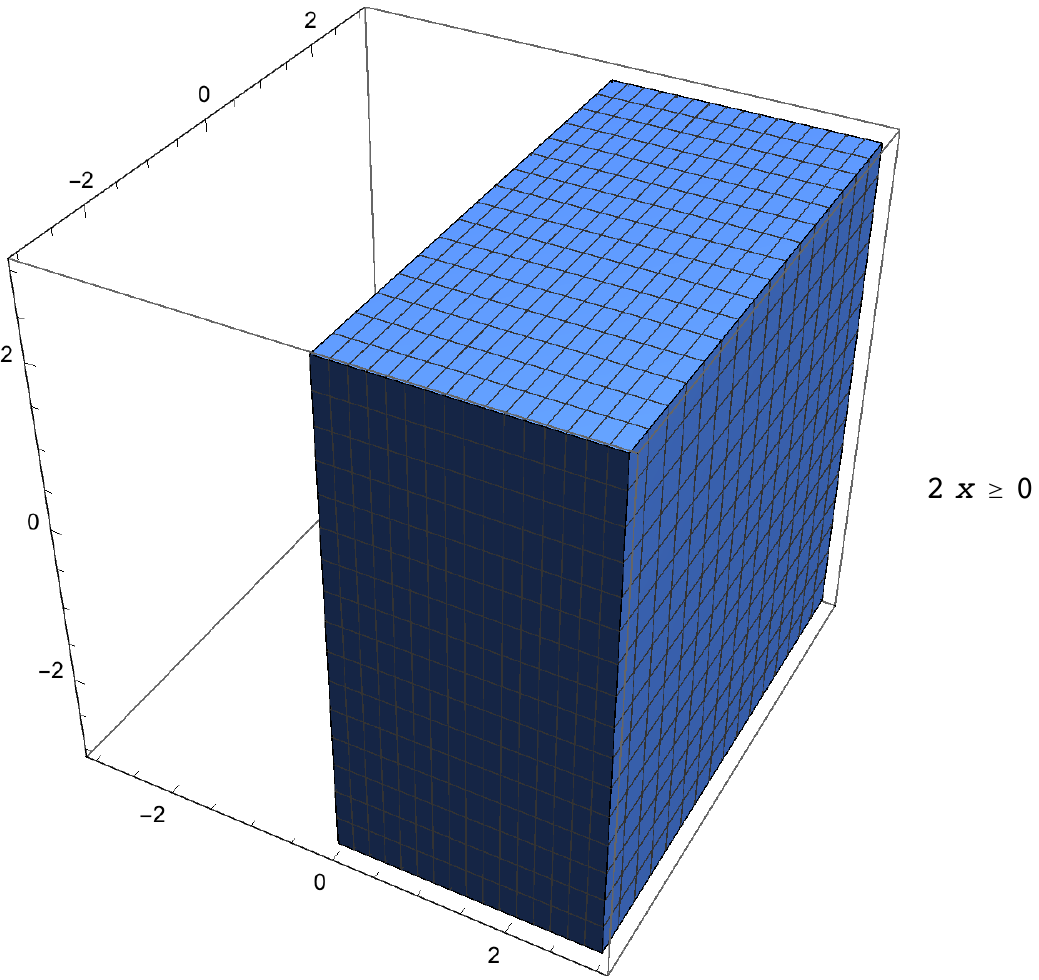}
 Input~8, $\cal S$
\end{minipage} 
\begin{minipage}{.19\textwidth}\centering
 \includegraphics[width=\textwidth]{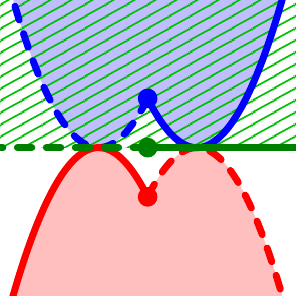}
 Input~9
\end{minipage} 
\caption{Interpolants from Table~\ref{tab:experimentResults}. The blue, orange and green areas are for $\mathcal{T}$,  $\mathcal{T}'$,  $\mathcal{S}$, respectively. }
\label{fig:experimentResults}
\end{figure}

\begin{figure}[p]
\begin{minipage}[t]{.3\textwidth}\centering\scriptsize
\begin{lstlisting}[label=codeDai11,caption={Code 1.3 of~\cite{DaiXZ13}}]
real x,y;
real xa = 0;
real ya = 0;
while(nondet()){
 x = xa + 2*ya;
 y = -2*xa + ya;
 x++;
 if(nondet()){
  y = y + x;
 }else{
  y = y - x;
 }
 xa = x - 2*y;
 ya = 2*x + y;
}
assert(xa + 2*ya >= 0);
\end{lstlisting}
\end{minipage}
\qquad\qquad
 \begin{minipage}[t]{.6\textwidth}\centering\scriptsize
 \begin{lstlisting}[label=codeAcc,caption=Constant Acceleration]
real x,v;
(x, v) = (0, 0);
while(nondet()){
 (x, v) = (x+2*v, v+2);
}
 assert(x >= 0);
 \end{lstlisting}

 \begin{align*}
 {\cal T} = \left(
\begin{array}{l}
(xa)+2(ya)\ge 0,  \enspace
 x=(xa)+2(ya),  \enspace
\\
y=-2(xa)+(ya), \enspace
x_1=x+1, \enspace
\\
y_1=x_1+y,\enspace 
(xa_1)=x_1-2y_1,\\
(ya_1)=2x_1+y_1
\end{array}
 \right)
\end{align*}

\vspace{-2em}
 \caption{The SAS for an execution of the code in Listing~\ref{codeDai11}}
 \label{fig:codeDai11SymbExec}
 \end{minipage}
\end{figure}

\vspace{.5em}
\noindent
\textbf{Program Verification Example I: Infeasibility Checking}\quad
 Consider the code in Listing~\ref{codeDai11}; this is from~\cite[\S{}7]{DaiXZ13}.
We shall solve  Subproblem~1 in~\cite[\S{}7]{DaiXZ13}:
 if the property
$(xa) + 2(ya) \ge 0$ holds at Line~5, then it holds too
after the execution along  $5\rightarrow 6\rightarrow
7\rightarrow 8 \rightarrow 9 \rightarrow 13 \rightarrow 14$.  The
execution is 
expressed as the  SAS $\mathcal{T}$ in Fig.~\ref{fig:codeDai11SymbExec}. 
Then our goal is to show that the negation 
\begin{math}
  {\cal T}' = \left((xa_1)+2(ya_1) < 0\right)
\end{math}
of the desired property is disjoint from $\mathcal{T}$. 


Our tool yields $\mathcal{S}=\bigl(8-14(ya_1)-7(xa_1)\ge 0\bigr)$ as an interpolant of these
$\mathcal{T}$ and $\mathcal{T}'$ (in 14.1 seconds, with parameters
$b=0, c=3$ and depth $d=11$). The interpolant witnesses disjointness. 
Our interpolant is far simpler than the interpolant given
in~\cite{DaiXZ13}.\footnote{An
interpolant $716.77+1326.74(ya)+1.33(ya)^2+
433.90(ya)^3+668.16(xa)-155.86(xa)(ya)+317.29(xa)(ya)^2+222.00(xa)^2+
592.39(xa)^2(ya)+271.11(xa)^3 > 0$ is given in~\cite{DaiXZ13}. 
We note that, to show disjointness of $\mathcal{T}$ and $\mathcal{T}'$, an interpolant of any splitting of $\mathcal{T} \cup \mathcal{T}'$ would suffice.  It is not specified~\cite{DaiXZ13} which splitting they used. 
}



Here the simplicity of our interpolant brings \emph{robustness} as its benefit. Consider the other path  $5\rightarrow \cdots \rightarrow 8 \rightarrow 11 \rightarrow
13 \rightarrow 14$ of execution from Line~$5$ to $14$, and let $\mathcal{T}_{0}$ be the SAS that expresses the execution. It turns out that our interpolant
 $\mathcal{S}$ in the above is at the same time an interpolant of $\mathcal{T}_{0}$  and $\mathcal{T}'$. Thus our algorithm has managed, aiming at simpler interpolants,  to automatically discover $-14(ya)-7(xa)$ (that is, $(xa) + 2(ya)$)
as a  value that is significant regardless of the choice made in Line~8.


\vspace{.5em}
\noindent
\textbf{Program Verification Example II: CEGAR}\quad
This is the example we discussed in
Example~\ref{ex:programverification}. Here we provide further details,
aiming at readers familiar with CEGAR. 

One of the most important applications of interpolation in verification is in
\emph{counterexample-guided abstraction
refinement (CEGAR)}~\cite{DBLP:journals/jacm/ClarkeGJLV03}. There an interpolant $\mathcal{S}$ is used as (a candidate for) the ``essential reason'' to distinguish positive examples $\mathcal{T}$ from negative counterexamples  $\mathcal{T}'$. 


As an example let us verify Listing~\ref{codeDai11} by CEGAR.
Starting from the empty set of abstraction predicates, CEGAR would find
the path $p_1 := (1\rightarrow 2 \rightarrow 3 \rightarrow 4 \rightarrow
16)$ as a counterexample.\footnote{Here we use a \emph{path-based} CEGAR
workflow that uses an execution path as a counterexample. Since we do not have any abstraction predicates, 
 $xa$ and $ya$ can be any integer; in this case the assertion in Line~16 can potentially fail.
} This counterexample path turns out to be spurious: let $\mathcal{T} := (xa=0,
ya=0)$ express the path and $\mathcal{T'} := ((xa) + 2 (ya) < 0)$ express the negation of the assertion; our tool $\SSPolyInt$ yields  $189346 (xa)+378692 (ya) \ge 0$ (i.e.\ $(xa)+2 (ya) \ge 0$) as an interpolant, proving their disjointness.
For the interpolation the tool $\SSPolyInt$ took $4.32$ seconds; we used the parameters $b=0$ and $c=5$.  

Consequently we add $(xa)+2 (ya) \ge 0$ as a new abstraction predicate and 
run the CEGAR loop again. This second run succeeds, since $(xa)+2 (ya) \ge 0$ turns out to be a suitable invariant for the loop in Line~4. We conclude safety of Listing~\ref{codeDai11}. 

We tried to do the same with the tool $\AISAT$~\cite{aiSatGitHub20170117,DaiXZ13} instead of our $\SSPolyInt$. It does not succeed in interpolating $\mathcal{T} = (xa=0,
ya=0)$ and $\mathcal{T'} = ((xa) + 2 (ya) < 0)$, since sharpness is required here (Prop.~\ref{prop:kojima}). As a workaround we tried strengthening 
$\mathcal{T'} = ((xa) + 2 (ya) < 0)$
into
$\mathcal{T}'_{0} = ((xa) + 2 (ya) \le 
-10^{-7})$;  $\AISAT$ then succeeded and yielded
an interpolant
$\mathcal{S}=(137.3430 +5493721088(ya)
+2746860544(xa) > 0)$.  
This predicate, however,  cannot exclude the
spurious path $p_1$ because
$\mathcal{S}$ and the negation $(xa) + 2 (ya) < 0$ of the assertion are satisfiable with 
 $xa = 0$ and
$ya = -1.25 \times 10^{-8}$.

\vspace{.5em}
\noindent
\textbf{Program Verification Example III: CEGAR}\quad
Here is another CEGAR example. Consider the code in
Listing~\ref{codeAcc} that models movement with constant acceleration. 
We initially have the empty set of abstraction predicates. After the first run of the CEGAR loop we would obtain a counterexample path
$p_1 := (1\rightarrow 2 \rightarrow 3\rightarrow 6)$; note that, since there are no predicates yet, $x$ can be anything and thus the assertion may fail. 

We let
 $\mathcal{T}_1 := (x=0,
v=0)$ express the counterexample $p_{1}$ and 
 $\mathcal{T}'_1 := (x < 0)$ express the negation of the assertion. 
For these $\mathcal{T}_1, \mathcal{T}'_1$ our tool $\SSPolyInt$ synthesizes $\mathcal{S}_1
:= (2x\ge 0)$ as their interpolant (in
$1.92$ seconds,  with $b=0$, $c=5$, and $d=1$).

Thus we add $2x\ge 0$ as an abstraction predicate and run the CEGAR loop again. 
We would then find the path $p_2 :=
(1\rightarrow  2\rightarrow 3\rightarrow 4
\rightarrow 5 \rightarrow 6)$ as a counterexample---note that the previous counterexample $p_{1}$ is successfully excluded by the new predicate  $2x\ge 0$. Much like before, we let $\mathcal{T}_2 := (v_1 = 0)$  express an initial segment of the path $p_{2}$ and let  $\mathcal{T}_2' := (x_1 = 0, v_2 =
v_1 + 2, x_2 = x_1 + 2 v_1, x_2 < 0)$ express the rest of the path $p_{2}$ (together with the negation of the assertion), and we shall look for their interpolant $\mathcal{S}_{2}$ as the witness of infeasibility of the path $p_{2}$. $\SSPolyInt$ succeeds, 
yielding $\mathcal{S}_2 := (8v_1 \ge 0)$ in $2.87$ seconds with $b =
0, c = 5, d = 1$.

In the third run of the CEGAR loop we use both $2x\ge 0$ and $8v \ge 0$ (from $\mathcal{S}_{1},\mathcal{S}_{2}$) as abstraction predicates. The proof then succeeds and we conclude safety of Listing~\ref{codeAcc}.

We did not succeed in doing the same with  $\AISAT$. In the first CEGAR loop
an interpolant of $\mathcal{T}_1$
and $\mathcal{T}_1'$ cannot be computed because it has to be sharp. 
As we did in the previous example  we could strengthen
$\mathcal{T}_1'$ to $\mathcal{T}_1'' := (x \le 10^{-7})$ and use an
interpolant of $\mathcal{T}_1$ and $\mathcal{T}_1''$ instead for the next iteration. 
$\AISAT$
generated an interpolant $3790.1050 +75802091520.0000 x > 0$ of $\mathcal{T}_1$ and $\mathcal{T}_1''$; however this fails to exclude the spurious counterexample path $p_{1}$. 

 Overall this example demonstrates that sharpness of interpolants can be a decisive issue in their application in program verification.

  \paragraph*{Acknowledgments}
 Thanks are due to
Eugenia Sironi, Gidon Ernst and the anonymous referees for their useful comments.
T.O., K.\ Kido and I.H.\ are supported by JST ERATO HASUO Metamathematics
for Systems Design Project (No. JPMJER1603), and JSPS Grants-in-Aid
No.\ 15KT0012 \& 15K11984.
K.\ Kojima is supported by JST CREST.
K.S.\ is supported by JST PRESTO No.\ JPMJPR15E5 and JSPS Grants-in-Aid No.\ 15KT0012. K.\ Kido is
supported by JSPS Grant-in-Aid for JSPS Research Fellows No.\ 15J05580.

\def\noopsort#1{} \def\singleletter#1{#1}

\newpage
\appendix

\section{Omitted Proofs}
\label{sec:proofs}

\subsection{Proof of Lem.~\ref{lem:ConeSOSRepr}}
\label{subsec:ConeSOSRepr}

\begin{mydefinition}[quadratic module]
  \label{def:quadratic module}
 A \emph{quadratic module} generated by $M=(m_\lambda)_{\lambda \in \Lambda} \in \real[\vec X]^\Lambda$ ($\Lambda$ is an index set) is the set
 \begin{align}
  \mathcal{QM}(M) \coloneqq
   \Set{q + \sum_{\lambda \in \Lambda}q_\lambda m_\lambda |
   \begin{aligned}
     & \text{$q\in \mathcal{C}(\emptyset)$,
       $(q_\lambda)_{\lambda\in \Lambda} \in \mathcal{C}(\emptyset)^\Lambda$, and}\\
     & \text{$q_\lambda = 0$ except for finitely many $\lambda$'s}
   \end{aligned}
   }.
 \end{align}
\end{mydefinition}

Lem.~\ref{lem:ConeSOSRepr} is an immediate consequence of the following lemma.

\begin{mylemma}
 \label{lem:coneqm}
 Let $f_1,\dots,f_s \in \real[\vec X]$.  Then,
 \begin{align}
  \mathcal{C}(f_1,\dots,f_s) 
  = \mathcal{QM}(f_1^{i_1}\dotsm f_s^{i_s} \mid i\in \mathbf{2}^s)
 \end{align}
 where $i=(i_1,\dots,i_s)$.
\end{mylemma}
\begin{myproof}
 The left-to-right direction is proved by induction on the
 construction of $\mathcal{C}(f_1,\dots,f_s)$.
 Let $f \in \mathcal{C}(f_1,\dots,f_s)$.
 \begin{enumerate}
  \item \underline{Case: $f=f_\lambda$ for some $\lambda=1,\dots,s$}.
	\begin{align}
	 f = f_\lambda
	 = f_1^0 \dotsm f_\lambda^1 \dotsm f_s^0
	 \in \mathcal{QM}(f_1^{i_1}\dotsm f_s^{i_s} \mid i\in \mathbf{2}^s).
	\end{align}
  \item \underline{Case: $f=g^2$ for some $g\in \real[\vec X]$}.
	\begin{align}
	 f = g^2
	 \in \mathcal{QM}(f_1^{i_1}\dotsm f_s^{i_s} \mid i\in \mathbf{2}^s).
	\end{align}
  \item \underline{Case: $f=g+h$ for some $g,h \in \mathcal{C}(f_1,\dots,f_s) 
	\cap \mathcal{QM}(f_1^{i_1}\dotsm f_s^{i_s} \mid i\in \mathbf{2}^s)$}.
      Obvious, because $\mathcal{QM}(f_1^{i_1}\dotsm f_s^{i_s} \mid i\in \mathbf{2}^s)$ is closed under addition.
  \item \underline{Case: $f=gh$ for some $g,h \in \mathcal{C}(f_1,\dots,f_s) 
	\cap \mathcal{QM}(f_1^{i_1}\dotsm f_s^{i_s} \mid i\in \mathbf{2}^s)$}.
	$g$ and $h$ have the form
	\begin{align}
	 g = \sum_{i\in \mathbf{2}^s} q_i f_1^{i_1} \dotsm f_s^{i_s},\quad
	 h = \sum_{j\in \mathbf{2}^s} r_j f_1^{j_1} \dotsm f_s^{j_s},
	\end{align}
	where $q_i, r_j \in \mathcal{C}(\emptyset)$.
        Let $p_\lambda = \sum_{i+j = \lambda} q_ir_j \in \mathcal{C}(\emptyset)$.
	Then
        \begin{align}
	 gh 
	 &=
	 \sum_{i\in \mathbf{2}^s}\sum_{j\in \mathbf{2}^s} q_i r_j f_1^{i_1+j_1} \dotsm f_s^{i_s+j_s}\\
	 &=
	 \sum_{\lambda\in \mathbf{2}^s} \sum_{i+j=\lambda} q_ir_j f_1^{\lambda_1}\dotsm f_s^{\lambda_s}\\
	 &= 
	 \sum_{\lambda\in \mathbf{2}^s} p_\lambda f_1^{\lambda_1}\dotsm f_s^{\lambda_s}\\
	 &=
	 \sum_{\lambda\in \mathbf{2}^s} \left(p_\lambda f_1^{2\lfloor \lambda_1/2 \rfloor}\dotsm f_s^{2\lfloor \lambda_s/2 \rfloor}\right) f_1^{\lambda_1 \bmod 2}\dotsm f_s^{\lambda_s \bmod 2}.
	\end{align}
	Because $p_\lambda f_1^{2\lfloor \lambda_1/2 \rfloor}\dotsm
        f_s^{2\lfloor \lambda_s/2 \rfloor} \in \cone(\emptyset)$ for each $\lambda \in 2^s$,
	we have $f=gh \in \mathcal{QM}(f_1^{i_1}\dotsm f_s^{i_s} \mid i\in \mathbf{2}^s)$.
 \end{enumerate}

 For the converse,
 let $f\in \mathcal{QM}(f_1^{i_1}\dotsm f_s^{i_s} \mid i\in \mathbf{2}^s)$.
 Then $f$ has the form
 $f = \sum_{i\in \mathbf{2}^s} q_i f_1^{i_1}\dotsm f_s^{i_s}$
 where $q_i \in \mathcal{C}(\emptyset) \subseteq \mathcal{C}(f_1,\dots,f_s)$.
 Then $f$ indeed belongs to $\cone(f_1,\dots,f_s)$, because
 cones are closed under addition and multiplication.
 \myqed
\end{myproof}

\subsection{Proof of Prop.~\ref{prop:kojima}}
\label{subsec:propKojima}

We argue by contradiction.  Assume that there exist polynomials $\tilde f, g,\tilde h$ 
 that satisfy the conditions in the proposition. 
 By the  feasibility assumption there exists some $\vec r\in \sem{\mathcal{T}_{\bullet}}\cap \sem{\mathcal{T}'_{\bullet}}$.
 For this $\vec r$ we have $\tilde f (\vec r) + \tilde h(\vec r) \ge 0$ by Lem.~\ref{lem:closurePropertiesOfCMI}, while by $1+\tilde f+ g^2+\tilde h=0$ we have 
$\tilde f (\vec r) + \tilde h(\vec r)= -1-\bigl(g(\vec r)\bigr)^{2} < 0$. Contradiction. 

\subsection{Proof of Thm.~\ref{thm:positivStellenSatzStrict}}
\label{subsec:ProofthmpositivStellenSatzStrict}
  The ``if'' direction is  easy. Suppose there exist such polynomials $f,g,h$ as required. Assume 
 $\vec r \in \sem{\mathcal{T}}$; then  we have
  $f(\vec r) \ge 0$, $g(\vec r) > 0$, and $h(\vec r) = 0$ by
  Lem.~\ref{lem:closurePropertiesOfCMI} and~\ref{lem:closurePropertiesOfS}.   Thus we have $f(\vec r) + g(\vec r) + h(\vec r) > 0$, which contradicts with $f + g + h = 0$.  Therefore
  $\sem{\mathcal{T}} = \emptyset$. 

  For the ``only if'' direction we rely on
  Thm.~\ref{thm:positivStellenSatz}, via the translation of
  \SASstrineq's to \SASdiseq's that we described after
  Def.~\ref{def:SAS2}, and use the fact that $\multMon(g_1, \dots, g_t)
  \subseteq \scone(g_1, \dots, g_t)$. \qed
  %

\subsection{Proof of Lem.~\ref{lem:SConeSOSRepr}}
\label{subsec:SConeSOSRepr}

We introduce a notion similar to quadratic module to represent strict cones.


\begin{mydefinition}[positive module]
  \label{def:positive module}
  An \emph{positive module} generated by $M=(m_\lambda)_{\lambda \in \Lambda} \in \real[\vec X]^\Lambda$ ($\Lambda$ is an index set) is the set
 \begin{align}
  \mathcal{PM}(M) \coloneqq
   \Set{r + \sum_{\lambda \in \Lambda}r_\lambda m_\lambda |
   \begin{aligned}
     & \text{$r \in \real_{\ge 0}$,
       $(r_\lambda)_{\lambda\in \Lambda} \in \real_{\ge 0}^\Lambda$,}\\
     & \text{$r_\lambda = 0$ except for finitely many $\lambda$'s, and} \\
     & \text{either $r_\lambda > 0$ for at least one $\lambda$, or $r > 0$}
   \end{aligned}
   }.
 \end{align}
\end{mydefinition}

Then Lem.~\ref{lem:SConeSOSRepr} is an immediate consequence of the
following lemma.
\begin{mylemma}
 \label{lem:sconeispm}
 Let $f_1,\dots,f_s \in \real[\vec X]$.  Then,
\begin{align}
 \mathcal{SC}(f_1,\dots,f_s)
 =
 \mathcal{PM}(f_1^{i_1}\dotsm f_s^{i_s} \mid i\in \nat^s).
\end{align}
\end{mylemma}
\begin{myproof}
  The left-to-right direction is proved by induction on the
  construction of $\scone(f_1, \dots, f_s)$.  Let $f\in \mathcal{SC}(f_1,\dots,f_s)$.
 \begin{enumerate}
  \item \underline{Case: $f=f_\lambda$ for some $\lambda=1,\dots,s$}.
	\begin{align}
	 f = f_\lambda = f_1^0 \dotsm f_\lambda^1 \dotsm f_s^0
	 \in \mathcal{PM}(f_1^{i_1}\dotsm f_s^{i_s} \mid i\in \nat^s).
	\end{align}
  \item \underline{Case: $f=r$ for some $r\in \real_{>0}$}.
	\begin{align}
	 f = r
	 \in \mathcal{PM}(f_1^{i_1}\dotsm f_s^{i_s} \mid i\in \nat^s).
	\end{align}
  \item \underline{Case: $f=g+h$ for some $g,h \in 
	\mathcal{SC}(f_1,\dots,f_s) \cap \mathcal{PM}(f_1^{i_1}\dotsm
        f_s^{i_s} \mid i\in \nat^s)$}.
      Obvious, because $\mathcal{PM}(f_1^{i_1}\dots f_s^{i_s} \mid i
      \in \nat^s)$ is closed under addition.
  \item \underline{Case: $f=gh$ for some $g,h \in 
	\mathcal{SC}(f_1,\dots,f_s) \cap \mathcal{PM}(f_1^{i_1}\dotsm f_s^{i_s} \mid i\in \nat^s)$}.
	$g$ and $h$ have the form of 
	\begin{align}
	 g = \sum_{i\in \nat^s} r_i f_1^{i_1} \dotsm f_s^{i_s},\quad
	 h = \sum_{j\in \nat^s} r'_j f_1^{j_1} \dotsm f_s^{j_s},
	\end{align}
        where $r_i, r'_j \in \real_{\ge 0}$.  Let
        $q_\lambda = \sum_{i+j = \lambda} r_ir'_j$.  Then obviously there are only finitely many $\lambda$ such that $q_\lambda > 0$.
        Moreover, there exists $\lambda$ such that $q_\lambda > 0$.
        Indeed, for $i$ and $j$ such that $r_i, r'_j > 0$, we have
	\begin{align}
          q_{i+j}
	 =
	 \sum_{i'+j'=i+j} r_{i'}r'_{j'}
	 \ge
	 r_{i}r'_j
	 >0.
	\end{align}
        Therefore,
	\begin{align}
	 gh
	 &=
	 \sum_{i\in \nat^s} \sum_{j\in \nat^s} r_i r'_j f_1^{i_1+j_1} \dotsm f_s^{i_s + j_s}\\
	 &=
	 \sum_{\lambda \in \nat^s} \sum_{i+j=\lambda} r_i r'_j f_1^{i_1+j_1} \dotsm f_s^{i_s + j_s}\\
	 &=
	 \sum_{\lambda\in \nat^s} q_\lambda f_1^{\lambda_1} \dotsm f_s^{\lambda_s}\\
	 & \in
	 \mathcal{PM}(f_1^{i_1}\dotsm f_s^{i_s} \mid i\in \nat^s).
	\end{align}

 \end{enumerate}
 For the converse, let us consider
 $f = r + \sum_i r_i f_1^{i_1}\dotsm f_s^{i_s} \in \mathcal{PM}(f_1^{i_1}\dotsm f_s^{i_s} \mid i\in \nat^s)$.
 For each $i$ such that $r_i > 0$, we have $r_i, f_1^{i_1},
 \dots, f_s^{i_s} \in \scone(f_1, \dots, f_s)$.  Moreover,
 either $r > 0$ or there exists at least one such $i$.  Therefore we conclude
 that $f \in \scone(f_1, \dots, f_s)$, because strict cones are
 closed under addition and multiplication.
 \myqed
\end{myproof}

\subsection{Proof of Lem.~\ref{lem:convergenceOfExtDiophantineApprox}}
\label{subsec:proofCFE}

We first notice that in case where the output of $\CFE(r, d-1)$ in
Line~10 equals $r$, then we can check that the value of $y$ at Line~11
equals $x$ by a straightforward calculation.

We can prove the lemma by induction on $\max\{ x_i \mid 1 \le i \le n\}$.
We check several cases in turn.
\begin{enumerate}
\item In the case where there is only one nonzero element in $x$,
  set $M = 1$.
  For simplicity, we consider the case $x = (k, 0, 0)$.
  Then $p = 1$ and $a = (1, 0, 0)$.
  Therefore if $d = 1$, then $y = a = x$ (as ratios).
  If $d > 1$, then $r = (k, 0, 0)$, so by induction on $d$ we obtain
  $r' = r$, and thus $y = x$ by the remark above.
\item In the case where there are more than one nonzero elements, and
  all of them are the same, set $M = 2$.
  For simplicity, we consider the case $x = (k, k, k, 0, 0)$.
  Then $p = 1$ and $a = (1, 1, 1, 0, 0)$.  If $d \ge M = 2$, then
  else-branch is taken, and $r = (k, 0, 0, 0, 0)$.
  By the previous case, we have $r' = r$.
  Therefore $y = x$ by the remark.
\item In other cases, let $r$ as in Line~9, and $M$ be the depth $d$
  for which $\CFE(r, d)$ stabilizes (such $M$ exists by the induction
  hypothesis).  Then for any $d \ge M + 1$ we have $r' = r$, and thus
  $y = x$ by the remark.
\end{enumerate}

The second claim is obvious from the construction of the algorithm.

\section{Topological  and Algebraic Closure}
\label{sec:closures}
Let us consider the difference between 
topological closure and algebraic closure, that is, the
difference between $\sem{\mathcal T_\bullet}\subseteq \real^{k}$ (where $\mathcal{T}_{\bullet}$ is from Def.~\ref{def:symbolicclosure}) and 
$\overline{\sem{\mathcal T}}\subseteq \real^{k}$. Here $\overline{(\place)}$ refers to the closure with respect to the Euclidean topology of $\real^{k}$.

They coincide in many cases but do not always.
For example, for $\mathcal T = (x^3-2x^2+x\le 0, x^3-2x^2+x\neq 0)$, 
we have $\sem{\mathcal T_\bullet} = \sem{x^3-2x^2+x\le 0} = (-\infty, 0] \cup \set{1}$,
but $\overline{\sem{\mathcal T}} = \overline{(-\infty, 0)} = (-\infty, 0]$.

We can show the following inclusion in general.
\begin{myproposition}
  \label{prop:whichbroader}
  Let $\mathcal A=(f_1 \triangleright_1 0, \dots, f_n \triangleright_n 0)$
  be an \SASdiseq, where $\triangleright_{i}\in\{\ge,\neq, =\}$. 
  Then
  \begin{align}
    \overline{\sem{\mathcal A}} \subseteq \sem{\mathcal A_\bullet}.
  \end{align}
\end{myproposition}
\begin{myproof}
 The proof is by  induction on $n$.
  \begin{itemize}
    \item \underline{Base cases}:  The cases $\triangleright_1 = (=)$ and
     $\triangleright_1 = (\ge)$ are easy because $f_1$ is continuous and 
     both $\set{0}$ and $[0,\infty)$ are closed in $\real$.
    For the remaining case where
 $\triangleright_1 = (\neq)$, we have
    $\overline{\sem{f_1 \neq 0}}\subseteq \real^k = \sem{()} = \sem{\mathcal A_\bullet}$.
    
    \item 
    \underline{Step case}: Let $\mathcal{B} = (f_1\triangleright_1 0,\dots, f_n\triangleright_n 0)$.
    \begin{align*}
      \overline{\sem{\mathcal A}}
      &=
      \overline{\sem{\mathcal B, f_{n+1}\triangleright_{n+1} 0}}\\
      &=
      \overline{\sem{\mathcal B} \cap \sem{f_{n+1}\triangleright_{n+1} 0}}\\
      &\subseteq
      \overline{\sem{\mathcal B}} \cap \overline{\sem{f_{n+1}\triangleright_{n+1} 0}} \\
      &\subseteq
      \sem{\mathcal B_\bullet} \cap \sem{(f_{n+1}\triangleright_{n+1} 0)_\bullet}
      \\
      &   
      \qquad \text{(by the induction hypothesis and arguments similar to the base case)}\\
      &=
      \sem{\mathcal A_\bullet}.
      \tag*{\myqed}
    \end{align*}
  \end{itemize}
\end{myproof}
It follows that 
$\overline{\sem{\mathcal T}} \cap \overline{\sem{\mathcal T'}} \subseteq \sem{\mathcal T_\bullet} \cap \sem{\mathcal T'_\bullet}$.
The opposite inclusion fails in general:
for $\mathcal{T} = (x^3-2x^2+x \le 0, x^3-2x^2+x \neq 0), 
\mathcal{T'} = (x\ge 0)$, $\overline{\sem{\mathcal T}}\cap \overline{\sem{\mathcal T'}}
= (-\infty, 0) \cap [0,\infty) = \emptyset$ and
$\sem{\mathcal T_\bullet}\cap\sem{\mathcal T'_\bullet} = 
((-\infty, 0) \cup \set{1}) \cap [0,\infty) = \set{1}$.

\section{Relationship of the Two Algorithms}
\label{sec:generalization}
We show that if Algorithm~\ref{alg:Dai} generates an interpolant for two \SASdiseq's, 
then Algorithm~\ref{alg:scone} generates an interpolant for two \SASstrineq's that are equivalent to those two 
\SASdiseq's.

\begin{myproposition}
  \label{prop:generalize}
  Let $\mathcal T$ and $\mathcal T'$ be the \SASdiseq's in (\ref{eq:T}).
  Let $\mathcal U$ and $\mathcal U'$ be the following \SASstrineq's.
  \begin{align}
    \footnotesize
    \begin{split}
      \mathcal{U}
    &=
    \left(
    \begin{array}{l}
      f_{1}(\vec{X},\vec Y) \ge 0\enspace,
     \;\dotsc,\;
     f_{s}(\vec{X},\vec Y) \ge 0\enspace,
     \quad
     g^2_{1}(\vec{X},\vec Y) > 0\enspace,
     \;\dotsc,\;
     g^2_{t}(\vec{X},\vec Y) > 0\enspace,
     \\
     h_{1}(\vec{X},\vec Y) = 0\enspace,
     \;\dotsc,\;
     h_{u}(\vec{X},\vec Y) = 0
    \end{array}
    \right),\\
    \mathcal{U}'
    &=
    \left(
    \begin{array}{l}
      f'_{1}(\vec{X},\vec Z) \ge 0\enspace,
     \;\dotsc,\;
     f'_{s'}(\vec{X},\vec Z) \ge 0\enspace,
     \quad
     {g'}^2_{1}(\vec{X},\vec Z) > 0\enspace,
     \;\dotsc,\;
     {g'}^2_{t'}(\vec{X},\vec Z) > 0\enspace,
     \\
     h'_{1}(\vec{X},\vec Z) = 0\enspace,
     \;\dotsc,\;
     h'_{u'}(\vec{X},\vec Z) = 0
    \end{array}
    \right).
    \end{split}
  \end{align}
  Obviously we have $\sem{\mathcal T} = \sem{\mathcal U}$ and $\sem{\mathcal T'}=\sem{\mathcal U'}$.

 Assume there exist polynomials
 $f,f',g,h,h'$ that satisfy the following conditions (cf.\ Thm.~\ref{thm:FromCertificateToInterpolant}). 
 \begin{itemize}
  \item  $f\in \cone(f_1,\dots,f_s), f'\in\cone(f'_1,\dots, f'_{s'}),
  g\in \multMon(g_1,\dots,g_t,g'_1,\dots,g'_{t'}), h\in \ideal(h_1,\dots,h_u)$,
  $h'\in \ideal(h'_1,\dots,h'_{u'})$, and
  \item $1+f+f'+g^2+h+h'=0$.
 \end{itemize}
Then
  there exist polynomials
  $\tilde f, \tilde f', \tilde g, \tilde h, \tilde h'$ that satisfy the following (cf.\ Thm.~\ref{thm:scone}). 
 \begin{itemize}
  \item  $\tilde f\in\cone(f_1,\dots,f_s,g_1^2,\dots,g_t^2),
  \tilde f' \in \cone(f'_1,\dots,f'_s,{g'}_1^2,\dots,{g'}_t^2),
  \tilde g\in \scone(g^2_1,\dots,g^2_t), \tilde h\in \ideal(h_1,\dots,h_u),
  \tilde h'\in \ideal(h'_1,\dots,h'_{u'})$ and
  \item $\tilde f+\tilde f'+\tilde g+\tilde h+\tilde h'=0$.
 \end{itemize}
\end{myproposition}
\begin{myproof}
  Set
  \begin{align}
    \tilde f=g^2+f\enspace,\quad \tilde f'=f'\enspace,\quad \tilde g = 1\enspace,\quad \tilde h = h\enspace,\quad \tilde h'=h'\enspace.
  \end{align}
  The equality $\tilde f+\tilde f'+\tilde g+\tilde h+\tilde h'=0$ is easy.
  Because $g^2$ is an SOS and $f\in \cone(f_1,\dots,f_s)$, $\tilde f\in \cone(f_1,\dots,f_s,g_1^2,\dots,g_t^2)$ hold.
  $\tilde f' \in \cone(f'_1,\dots,f'_s,{g'}_1^2,\dots,{g'}_t^2)$ is obvious.
  By the definition of strict cone, $\tilde g = 1 \in \scone(g_1,\dots,g_t)$. 
  $\tilde h\in \ideal(h_1,\dots,h_u),
  \tilde h'\in \ideal(h'_1,\dots,h'_{u'})$ are obvious.
  \myqed
\end{myproof}

Recall that Algorithm~\ref{alg:Dai} generates an interpolant based on Thm.~\ref{thm:FromCertificateToInterpolant}, and that Algorithm~\ref{alg:scone} is based on Thm.~\ref{thm:scone}. Therefore by Prop.~\ref{prop:generalize}, if the former succeeds,  so does the latter.

\end{document}